\documentclass[useAMS,usenatbib]{mn2e}

\usepackage{amsmath,graphicx,wrapfig,lipsum,caption,overpic,amssymb,cprotect}

\newcommand{\ignore}[1]{}

\def\mdot{\dot{M}}
\title[Composition of Early Planetary Atmospheres I]{Composition of Early Planetary Atmospheres I: Connecting Disk Astrochemistry to the Formation of Planetary Atmospheres}
\author[A. J. Cridland, R. E. Pudritz and M. Alessi]{A. J. Cridland$^{1}$\thanks{E-mail:
cridlaaj@mcmaster.ca}, R. E. Pudritz$^{1,2,3,4}$\thanks{E-mail:
pudritz@mcmaster.ca} and M. Alessi$^{1}$\thanks{E-mail:
alessimj@mcmaster.ca}\\
$^{1}$Department of Physics and Astronomy, McMaster University, Hamilton, Ontario, Canada, L8S 4E8 \\ $^2$Origins Institute, McMaster University, Hamilton, Ontario, Canada, L8S 438 \\ $^3$Zentrum f¨ur Astronomie der Universit¨at Heidelberg, Institut f¨ur Theoretische Astrophysik, Albert-Ueberle-Str. 2, 69120 Heidelberg, Germany \\ $^4$Max Planck Institute for Astronomy K¨onigstuhl 17, D-69117 Heidelberg, Germany}

\begin{document}

\date{\today}

\pagerange{\pageref{firstpage}--\pageref{lastpage}} \pubyear{2015}

\maketitle

\label{firstpage}

\begin{abstract}
We present a model of the early chemical composition and elemental abundances of planetary atmospheres based on the cumulative gaseous chemical species that are accreted onto planets forming by core accretion from evolving protoplanetary disks. The astrochemistry of the host disk is computed using an ionization driven, non-equilibrium chemistry network within viscously evolving disk models. We accrete gas giant planets whose orbital evolution is controlled by planet traps using the standard core accretion model and track the chemical composition of the material that is accreted onto the protoplanet. We choose a fiducial disk model and evolve planets in 3 traps - water ice line, dead zone and heat transition. For a disk with a lifetime of $4.1$ Myr we produce two Hot Jupiters (M $= 1.43, 2.67$ M$_{\rm Jupiter}$, r $= 0.15, 0.11$ AU) in the heat transition and ice line trap and one failed core (M $= 0.003$ M$_{\rm Jupiter}$, r $=3.7$ AU) in the dead zone. These planets are found with mixing ratios for CO and H$_2$O of $1.99\times 10^{-4}$, $5.0\times 10^{-4}$ respectively for both Hot Jupiters. Additionally for these planets we find CO$_2$ and CH$_4$, with mixing ratios of $1.8\times 10^{-6}\rightarrow 9.8\times 10^{-10}$ and $1.1\times 10^{-8}\rightarrow 2.3\times 10^{-10}$ respectively. These ranges correspond well with the mixing ratio ranges that have been inferred through the detection of emission spectra from Hot Jupiters by multiple authors. We compute a carbon-to-oxygen ratio of $0.227$ for the ice line planet and $0.279$ for the heat transition planet. These planets accreted their gas inside the ice line, hence the sub-solar C/O.
\end{abstract}


\section{ Introduction }\label{sec:intro}

Understanding the chemical composition and structure of exoplanetary atmospheres is now feasible with the advent of large surveys from the Hubble space telescope and Lowell observatory (TrES - Transatlantic Exoplanet Survey, \cite{Alo04}) along with planned missions like the Transiting Exoplanetary Survey Satellite (TESS), Extremely Large Telescope (ELT) follow up to PLATO, and the James Webb Space Telescope (JWST). A number of authors (\cite{MadSea09,MadSea10}, \cite{Lee11,Lee13}, \cite{Ven14}, \cite{Lin11,Lin13}, \cite{Stv10}, \cite{MigKal13}, \cite{VGro14},\cite{Sing14}, \cite{Kre14}, \cite{Brog16},\cite{Kat16},\cite{Okl16}) have begun retrieving chemical abundance data from the spectra of directly observed, and transiting exoplanets. These efforts have detected the chemical signatures of about a dozen chemical species including H$_2$O, CO, CO$_2$, CH$_4$, HCN, Na and Mg for about a dozen planets.

 By an early atmopshere we mean a young system ($t_{age} \lesssim 10$ Myr) that has not seen enough dynamical and chemical evolution to alter the chemical composition of the gas that it accretes from its natal protoplanetary disk. Computing the early chemical composition of an exoplanetary atmosphere requires the combination of three areas of study: accretion disk physics, astrochemistry, and planet formation physics. The work presented here combines these studies in a series of steps. The local chemical composition of a disk at any position or time is indicated by non-equilibrium chemical processes in the viscously evolving disk. Planetary cores migrating through their host disk will accrete materials of different chemical composition as they move. The final mix of materials accreted onto the planet will therefore depend on the astrochemistry of the evolving disk, and the details of its accretion history during its formation.

 In past works, the astrochemistry of accretion disks has been studied by applying numerical, 1+1D photo-chemical codes as seen in \cite{Berg03}, \cite{Fog11} and \cite{Cle14} and 2D radiative thermo-chemical codes (ie. ProDiMo, see \cite{Wot09}) in the context of static, viscous and radiatively heated accretion disks (eg. \cite{Dal98,Dal99}, \cite{LB74} and \cite{CG97}). For a comprehensive review of this topic see \cite{SeHe13}. Population synthesis models as given in \cite{IL04} and \cite{HP13} have been used to study the physics of planet formation. These population synthesis models rely on the core accretion model \citep{KI02} which describes the rate of formation of a protoplanet out of a sea of planetessimals ranging between 1-10 km in size. 

It is the goal of this paper to combine these streams of research to compute the chemical composition of the gas that is delivered to planets. This provides the initial conditions for building models of planet and atmospheric structure that do not depend on the assumption of solar abundances. Indeed we will show that there are significant deviations from solar abundances depending on {\it where} and {\it when} the planet forms in the disk. 

For example the carbon and oxygen content of an exoplanetary atmosphere has been compared to the composition of gas at the location where the protoplanet accretes its gas (see for example \cite{Mad14}). To first order, the chemical content of the atmosphere is set by the location where the planet accretes its gas relative to the condensation fronts of H$_2$O, CO$_2$ and CO. At larger radii than the condensation front a particular chemical species resides in the solid (frozen) phase, while at smaller radii it is a gas. As shown in \cite{Mad14}, tracking the mass and orbital evolution of the accreting planet allows for the determination of the chemical content of planets at different final semi-major axes.

A more complete analysis of the connection between disk astrochemistry and planetary atmospheres was done by \cite{Hel14}. In this work the chemical structure of the material available for planet formation was computed using the ProDiMo code \citep{Wot09} in a  \cite{LB74} disk model. \cite{Hel14} showed that after a short relaxation time ($<< 10^3$ yr) the gas chemistry reaches an equilibrium state at the midplane of the disk, where the time evolution becomes dominated by cosmic ray ionization driven reactions. This time evolution is traced by the elemental ratio of carbon to oxygen (C/O) in the gas, which increases with time in their model. Their connection to planetary atmospheres was made by studying the formation of clouds.  Cloud formation is sensitive to the presence of silicate grains in the atmosphere that acts as a seed for clouds. \cite{Hel14} demonstrated that oxygen poor planets, with C/O $\sim 1$, formed clouds less efficiently than oxygen rich planets because of the lack of oxygen to form 
silicates.

\ignore{The growing planet in our model forms in an evolving isolated disk, as material is either accreted through the disk and onto the host star by disk viscosity and winds, or is striped away by high energy photons.}

Absent from previous works is the important fact that material is accreted through the disks and onto the host star due to torques caused by disk winds and viscosity, and is later photoevaporated by FUV photons. This accretion history changes the surface density and global temperature structure of the disk as it ages, thereby changing the location of the condensation fronts while the planets evolve. This evolution of disk accretion is also a driver for the evolution of the chemistry within disk. Finally, photoevaporation ultimately sets the lifetime of disks, which determines the planet's final mass and ends their orbital evolution by purely disk-planet interactions. Our method of modeling these important physical processes will be discussed below.

The lifetime of a disk is set by the strength of photoevaporation from the host star, and is between $3-4$ Myr on average (eg. \cite{Her07}, \cite{HLL01}, \cite{Strom89} and \cite{Wilk89}). The source of the photoevaporating UV rays is the hot accretion shock ($\sim 9000$ K) produced by material accreted onto the star along magnetic field lines \citep{Gort15}. The accretion shock also acts as a point source of X-rays which are partially responsible 
for setting the ionization structure of the disk. The ionization structure controls the coupling between the magnetic field and the gas disk (an important feature of MRI driven turbulence, see below) as well as the production of ions. These ions are an important part of gas phase chemistry because of their low activation barrier during reactions with neutral molecules.

\begin{figure}
\centering
\includegraphics[width=0.48\textwidth]{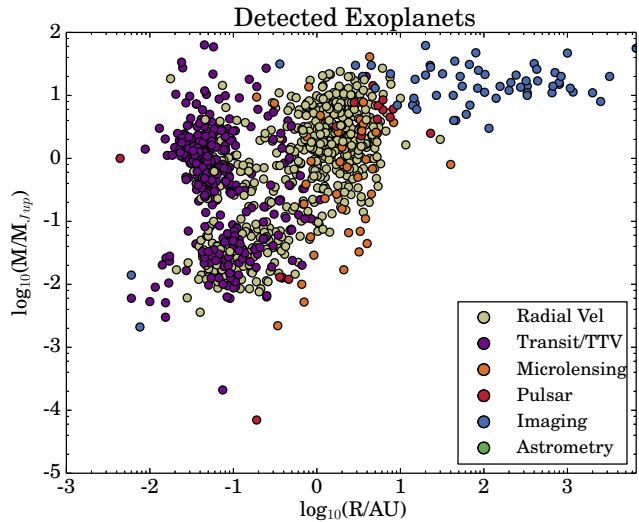}
\caption{All available exoplanetary data accessed from http://exoplanet.eu/catalog/ on November 23rd, 2015. The colour of the scatter points show the detection method used to determine the mass of the planet.}
\label{fig:intro01}
\end{figure}

Apart from these factors that impact isolated disks, several external agents can change how the structure of the disk evolves when the young system is in a cluster environment. These include dynamical interactions between stars that could truncate disks, and the influence of the cluster UV and X-ray background arising from O and B stars in the cluster. However, observational studies have shown that the proximity to the most massive stars in a cluster does not affect the survivability of disks \citep{R15}. Furthermore dynamical interactions with other cluster stars do not scatter planets with orbital separations less than $\sim 200$ AU from their host star \citep{Z15}. Most observed planets have orbital separations much less than $\sim 200$ AU (see Figure \ref{fig:intro01}), so we choose to ignore the impact of these external effects.

Gas chemistry in the disk is dominated by ion-molecule, photodissociation and dissociative recombination reactions. These reactions are supplemented by neutral-neutral molecular reactions, however they are limited by an activation barrier and generally have reaction rates that are two orders of magnitude smaller than the above reactions  \citep{Tie05}. On the surface of dust grains, the reaction rates of these neutral-neutral  reactions are increased because the collisional probability of two gas molecules is higher than in free space. As a result, grain surface reactions are very important for the production of more complicated gas species like formaldehyde (H$_2$CO) which is formed through the proton capture of frozen carbon monoxide. As discussed above there is a stellar source for the high energy photons responsible for the photodissociation and ionizing reactions in the gas phase, however these photons may not reach deep into the interior of the disk. 
A second source of photodissociating photons can be produced from the radiative relaxation of excited H$_2$ molecules. The source of these excited molecules is from primary electrons produced by cosmic rays \citep{Tie05}.

Astrochemical models of static protoplanetary disks are well developed \citep{Wot09,SW11,Fog11,SeHe13,Thi13,Wal13,Cle14}. These combine passive models (no time evolution) of 2D protoplanetary disks with advanced methods of high energy radiative transfer (eg. \cite{BB11x}). These models are combined with a large set of gas phase reactions that number over 6000 (eg. the destruction of N$_2$H+ by carbon monoxide gas: Equation \ref{rct:01}) with grain surface reactions (eg. the formation of water by proton capture of frozen oxygen: Equation \ref{rct:02}) and photodissociative reactions (eg. Equation \ref{rct:03}) to produce a complete snapshot of the chemical structure in the disk. 

\begin{align}
{\rm N}_2{\rm H}^+ + {\rm CO} &\rightarrow {\rm HCO}^+ + {\rm N}_2 \label{rct:01}\\
{\rm H} + {\rm H} + {\rm O(gr)} &\rightarrow {\rm H}_2{\rm O(gr)} \quad {\rm (net)} \label{rct:02}\\
{\rm HCO} + \gamma &\rightarrow {\rm H} + {\rm CO}\label{rct:03}
\end{align}

A significant range of molecules has been detected by large sub-millimeter arrays such as ALMA, IRAM and SMA. At the time of writing, emission lines from H$_2$O, CO, HCN, NH$_3$, C$_2$H$_2$, CO$_2$ and OH \citep{Pon14} have been detected, including many that are important to the origins of life (eg. Glycolaldehyde \citep{J12}). Even with this wealth of data the full content of carbon and nitrogen based molecules in accretion disks has not been fully determined \citep{Pon14}. 

We now turn to the second aspect of the problem: accretion and migration processes involved in planet formation. We adopt the core accretion picture of planet formation within the inner 100 AU of protoplanetary disks, which has broad observational support \citep{IL04,Mord12,And13,ILN13,HP13,D14}. This model is appropriate for the majority of gas giant planets which have been observed within a few AU, and the new population of planets known as Hot Jupiters that orbit at semi-major axes less than a tenth of an AU (see Figure \ref{fig:intro01}). 

Successful methods of finding exoplanets include radial velocity, transit methods and direct imaging. The majority of planets have been reported in the last five years \citep{Kep12,Kep14}. Inferring the chemical abundance of an exoplanetary atmosphere requires the detection of the transmission or emission spectra from transiting or nearly transiting exoplanets (eg \cite{Lee11,Lee13,Snel15}) or direct measurement of the emission spectra for a direct imaging planet \citep{Kon14}. The most abundant molecules found so far are: H$_2$O, CO, CO$_2$ and CH$_4$, along with some trace detections of Na and K \citep{Sing14,Wil15,Wyt15,Mont15}. There remains a wide range of uncertainty in these detections, mostly caused by systematic errors in observational techniques \citep{Crl15}. Furthermore there are insensitivities in the resulting features of observed emission spectra from chemical atmospheric models that vary chemical element ratios (like C/O, see \cite{Mos12}). 

Moreover, chemistry impacts the physical system through the dependence of the planetary atmosphere's total opacity at a given wavelength to the chemical abundance of the gas \citep{SD10}. The opacity of the gas sets its temperature and temperature gradient, thereby affecting the way that energy can be transported through the atmosphere. When the opacity is very high, energy will be transferred through convection because the radiative temperature gradient in the planetary atmosphere becomes very large. Having a good estimate of the gas opacity is dependent on having an accurate account of present chemical species in the gas. 

Our astrochemistry model follows the same methods used in \cite{Fog11} and \cite{Cle14}, however we follow the evolution of the chemical structure by computing the physical and chemical state of the disk over 300 snapshots extending over 4 million years.

The paper is organized as follows: our disk model and chemical model are outlined in \S 2 and \S 3 respectively. The migration and accretion model is discussed in \S 4. We compute and report initial chemical compositions of early atmospheres in \S 5. We compare our results to observations and discuss their implications in \S 6. Concluding remarks are found in \S 7.

\graphicspath{{../../Research/CASCA15/}}

\section{ Disk Model }\label{sec:meth}

Computing the astrochemistry in an accretion disk requires the temperature and surface density structure, as well as the local distribution of ionizing photons.  Accretion disks are heated by a combination of viscosity in the inner regions \citep{LB74} and radiative heating \citep{CG97}. Both of these heating sources produce power laws that describe the temperature and surface density profile, with exponents that depend on which heating source is dominating. It has been demonstrated through numerical work that the true midplane temperature profile can be described by a double power law. The inner region (within the heat transition point) mimics the temperature profile caused by a viscously heated disk. The outer region of the disk mimics the profile caused by direct illumination \citep{Dal98,Dal99}. From this numerical picture it has become clear that a purely viscous disk would be too cold in the outer regions, affecting the distribution of important ices like CO and NH$_3$. A purely radiative disk on the 
other hand would be too cool at smaller radii, affecting the distribution of important gases like H$_2$O and CO$_2$. 

In our work, we wish to incorporate the double power law form of the temperature and surface density profiles observed in numerical simulations. However, a full numerical treatment of the disk is too computationally expensive when the disk properties must be traced over millions of years of evolution. Fortunately, a model introduced by \cite{Cham09} reproduces the disk structure caused by two separate heating sources while remaining fully analytic, allowing for the density and temperature structure to be traced over the whole life of the disk. 

The temperature and surface density is described by a power law,
\begin{align}
T\left(r\right) \propto r^{-\beta}  {\rm ;}\quad \Sigma\left(r\right) \propto r^{-s},
\label{eqn01}
\end{align}
where the exponents $\beta$ and $s$ are given by the source of heating in the disk. These exponents are determined by solving the diffusion equation for a viscously evolving disk \citep{LB74},
\begin{align}
\frac{\partial \Sigma}{\partial t} = \frac{3}{r} \frac{\partial}{\partial t} \left[ r^{1/2} \frac{\partial}{\partial t} \left(r^{1/2} \nu\Sigma\right)\right],
\label{eq:in01}
\end{align}
based on the previous work of \cite{St98} which included heating from both viscous evolution and direct irradiation. The \cite{Cham09} solution is self-similar, with the variable $r/s(t)$ where $s(t)$ is the time-dependent size of the disk. This approach computes the temporal evolution of the temperature and surface density profiles simultaneously, and relies on how the viscosity of the disk varies, as a power-law in radius \citep{LB74}. 

Angular momentum transport that enables disk accretion can be driven radially by turbulent stress, or vertically due to the MHD torque exerted by a magnetized disk wind (eg. see \cite{PPV07,BS13,Gres15a}). We assume that both the effective viscosity due to either turbulent or wind stress $\nu$ can be modeled following the standard $\alpha$-disk model \citep{SS73}\begin{align}
\nu = \alpha c_s H,
\label{eq:in01b}
\end{align}
where $c_s$ is the sound speed of the gas and $H$ is the pressure scale height. In Equation \ref{eq:in01b}, the $\alpha$ parameter for turbulent transport is generally defined as the ratio between the turbulent and thermal pressures \citep{SS73}. While in the case of angular momentum loss to the disk by disk winds, it is the ratio between the vertically directed component of the Maxwell stress and the thermal pressure. Generally speaking the disk's effective $\alpha$ is due to a combination of both of these transport mechanisms: $\alpha \equiv \alpha_{turb} + \alpha_{wind}$.

In this work we assume that the source of the turbulence is caused by the operation of a magnetorotational instability (MRI) \citep{BH91}. The strength of the MRI driven turbulence (magnitude of $\alpha_{turb}$) depends on the level of ionization in the disk. It is characterized by the Ohmic Elsasser number, which is the ratio of the dissipation timescale to the growth timescale of the most unstable mode:
\begin{align}
\Lambda_O \equiv \frac{\tau_{diss}}{\tau_{grow}} = \frac{v_{A,z}^2}{\eta_O \Omega}.
\label{eq:in02}
\end{align}
In this expression $v_{A,z} \equiv B_z/\sqrt{4\pi \rho}$ is the Alfv\'en speed in the z-direction, $B_z$ is the local magnetic field in the z-direction, $\rho$ is the gas density and $\Omega$ is the orbital frequency \citep{BS13,Gres15a}. When $\Lambda_0 > \Lambda_{0,crit} \equiv 1$ \citep{Simon13} we assume that that region of the disk is turbulently active, with $\alpha_{turb} = 10^{-3}$, while in the opposite case we assume that the disk is turbulently `dead' with $\alpha_{turb} = 10^{-5}$. The Ohmic resistivity,
\begin{align}
\eta_O = \frac{234 ~T^{1/2}}{x_e} ~cm^2s^{-1},
\label{eq:in03}
\end{align}
depends on the electron fraction ($x_e$) and temperature structure in the disk \citep{KB04}. 

The Ohmic resistivity will vary throughout the disk due to the changing temperature and ionization structure of the gas. This leads to changes in the activity of the MRI. Our research differs from previous works (see \cite{SHT09} and \cite{HP13}), where a semi-analytic approach was taken to compute the electron fraction. Instead, we compute the electron fraction of the disk directly with our photochemical code for a given temperature and surface density profile to determine the Ohmic Elsasser number. At the midplane of the disk, MRI-inactive regions will be present where there is low ionization. This has implications on the global angular momentum transport and ultimately on the mass accretion through the disk. 

It has been suggested that angular momentum transport can be maintained through the dead zone by other means such as magnetocentrifugally driven winds \citep{PN83} or layered accretion through regions higher in the disk's atmosphere \citep{LS11}. In both cases, the mass is accreted through higher regions of the disk where densities are lower and more susceptible to angular momentum transport by MRI driven turbulence or winds \citep{Gres15a,Gres15b}. Because the mass accretion rate through dead zones can be maintained by these mechanisms, we assume that the mass accretion rate $\mdot$ remains constant in radius through the disk. As a mathematical constraint, we assume that the disk effective viscosity parameter $\alpha \equiv \alpha_{turb} + \alpha_{wind}$ also remains constant in both space and time. 

In light of the idea of a turbulently dead zone, we will treat the chemical structure of the disk as a passive feature of our disk model. Any impact on the physical structure from the chemical evolution (eg. opacity, $\alpha$ or density jumps at condensation fronts) is thus ignored. The inclusion of the effect of the chemistry can be achieved through an iterative process as was done for brown dwarf and hot Jupiter atmospheres \citep{Mar96,For05}. However this iterative process is computationally expensive and requires multiple calculations of the temperature and chemical structure in the disk, with no guarantee of a converging solution.

\ignore{
Some recent work has suggested that a dead zone may remain hydrodynamically unstable on large scales, generating turbulence due to the Zombie Vortex Instability (ZVI) \citep{Mac15}. Numerical simulations have shown that the ZVI could produce turbulent strengths rivaling the MRI active regions, however numerical resolution and setup have limited the interpretation of these results \citep{Mac15}.}

As mass is accreted onto the host star due to angular momentum transport by viscous stress, the disk must spread out, which forces a reduction in mass accretion rate. In the \cite{Cham09} model, the mass accretion rate decreases as a power law in time (see Appendix \ref{sec:calcmod}). It has been shown in population synthesis models however \citep{HP13} that a dissipative model of the mass accretion rate more accurately reproduces the population of exoplanets that have been observed. The photoevaporation of disks results in an exponential decay of the mass accretion rate, so that:
\begin{align}
\mdot \equiv \mdot_{vis}(t) \rightarrow \mdot\equiv \mdot_{vis}(t)e^{-(t-t_{in})/t_{dep}},
\label{eq:mdot01}
\end{align}
where $\mdot_{vis}$ is the equilibrium accretion rate determined by the solution to the viscous Equation \ref{eq:in01}, $t_{in}$ is the starting time of the simulation (0.1 Myr), and $t_{dep}$ is an estimate of the depletion time of the dissipation. The standard source of the dissipation is assumed to be the photoevaporation \citep{HP13} of the disk from the UV or X-ray emission of the host star. The impact of photoevaporation on disk structure has been studied in great detail, and impacts both the overall surface density structure and lifetime of accretion disks around young stars \citep{Alex14}. It has been shown \citep{Gort15} that disk dispersal due to photoevaporation happens very quickly, with time scales on the order of $10^4$ years. Because of this small timescale we treat the disk lifetime as a free parameter, constrained by observations, and assume that at $t=t_{life}$ the 
disk is abruptly cleared. The full form of equation \ref{eq:mdot01} is \begin{align}
\mdot\equiv \begin{cases}
    \mdot_{vis}(t)e^{-(t-t_{in})/t_{dep}} & t < t_{life}\\
    0 & t \ge t_{life}.
    \end{cases}
\label{eq:mdot02}
\end{align}
Any planets that are forming in the disk at $t_{life}$ are frozen in orbital radius when the disk is cleared, and we assume no further orbital or mass evolution takes place due to gas dynamical processes. The values used for $t_{in}$, $t_{dep}$ and $t_{life}$ are found in Table \ref{tab:result01} below.

\subsection{ Heating Sources }

The sources of heating determine the surface density and temperature distributions, as well as how the distributions depend on the evolution of the mass accretion rate with time. Ultimately, the heating is also strongly connected to the chemical makeup of the disk, through the temperature dependence of most chemical reactions.

\subsubsection{ Viscous Heating }\label{sec:VisHeat}

In the viscous regime, heating results from the viscous stress on the disk. The resulting effective temperature, $T_{eff}$,  describes the flux of radiation emitted by a ring of radius R caused by the viscous accretion of material through the ring, assuming the material acts as a blackbody. The effective temperature is given by \citep{Cham09},
\begin{align}
2\sigma T_{eff}^4 = \frac{9\nu\Sigma\Omega^2}{4}.
\label{eqn02}
\end{align}
The temperature at the midplane is then:
\begin{align}
T^4 = \frac{3\tau}{4} T_{eff}^4,
\label{eqn03}
\end{align}
where the optical depth is defined by:
\begin{align}
\tau = \frac{\kappa_0 \Sigma}{2},
\label{eqn04}
\end{align}
where $\kappa_0 = 3 ~cm^2/g$ is an assumed constant opacity.

With that, we express the midplane temperature as 
\begin{align}
\left(\frac{T}{T_{vis}}\right)^4 = \left(\frac{\nu}{\nu_0}\right)\left(\frac{\Sigma}{\Sigma_0}\right)^2\left(\frac{\Omega}{\Omega_0}\right)^2.
\label{eqn05}
\end{align}
All subscripts of $0$ denote the value of all variables at the outer edge of the disk at $t=0$, denoted by $s_0$ and 
\begin{align}
T_{vis}^4 = \left(\frac{27 \kappa_0}{64 \sigma}\right) \nu_0\Sigma_0^2\Omega_0^2.
\label{eqn06}
\end{align}

It has been shown that accreting disks enter into a quasi-steady state \citep{PT99} with a constant (in radius) mass accretion rate of $\mdot = 3\pi\nu\Sigma$ \citep{Alb05,Cham09}. In more recent simulations (see \cite{BC14} and \cite{BCP15}), this result has been confirmed at later disk lifetimes ($t > 10^5$ yr) for radii less than a few tens of AU. Following these previous works we will assume that the mass accretion rate is constant in radius over the whole disk, with the value of the steady state solution from above. Due to angular momentum conservation one would require that
\begin{align}
\mdot_{vis} &= 3\pi \nu\Sigma &r < r_{turn} \nonumber\\
\mdot_{vis} &= -3\pi \nu\Sigma  &r > r_{turn},
\label{eqn10}
\end{align}
where $\mdot_{vis}$ is the same mass accretion rate shown in Equation \ref{eq:mdot01} and \ref{eq:mdot02}, and $r_{turn}$ is the hypothetical turning point of the gas accretion direction. For visual simplicity we drop the subscript from the mass accretion rate in Equation \ref{eqn10} and assume in the final expressions that the mass accretion rate is determined by both viscous (Eq. \ref{eqn10}) and evaporative (Eq. \ref{eq:mdot02}) sources. Viscous heating only depends on the magnitude of the accretion rate, not the direction of accretion. In what follows, only the magnitude of $\mdot$ will be important and not its sign. Combining equation \ref{eqn05} with the mass accretion rate results in the radial dependence of the surface density:
\begin{align}
\Sigma = \Sigma_{vis}\left(\frac{\mdot}{\mdot_0}\right)^{3/5}\left(\frac{r}{s_0}\right)^{-3/5},
\label{eqn11}
\end{align}
this results in the temperature profile: \begin{align}
T = T_0 \left(\frac{\mdot}{\mdot_0}\right)^{2/5}\left(\frac{r}{s_0}\right)^{-9/10},
\label{eqn11x}
\end{align}
where the initial mass accretion rate $\mdot_0 = 3 \pi \nu_0\Sigma_0$ is set by the choice of initial surface density $\Sigma_0$ at the edge of the disk, which for a purely viscous disk is exactly $\Sigma_0 = \Sigma_{vis}$. This density can be constrained by computing the total mass,
\begin{align}
M \equiv 2\pi \int^s_0 r\Sigma dr,
\label{eqn12}
\end{align}
at the initial time. Substituting in Equation \ref{eqn11} we find $\Sigma_{vis} = 7M_0/10\pi s_0^2$, where $M_0$ is the initial mass of the disk. Using conservation of angular momentum one finds (see \cite{Cham09}) that the size of the disk scales with the total mass as:
\begin{align}
\left(\frac{s}{s_0}\right) = \left(\frac{M}{M_0}\right)^{-2}.
\label{eqn13}
\end{align} 
The spread in the radial direction is primarily due to the viscous evolution of the fluid rather than other angular momentum transport mechanisms like winds, which direct angular momentum vertically. So we assume that the spreading rate will only depend on the mass accretion rate due to viscous evolution. In combining eqs. \ref{eqn10}, \ref{eqn11}, \ref{eqn12} and \ref{eqn13} we find:
\begin{align}
\frac{dM}{dt} \equiv -\mdot_{vis} = -\mdot_0 \left(\frac{M}{M_0}\right)^{19/3}.
\label{eqn14}
\end{align}

\subsubsection{ Radiative Heating }\label{sec:RadHeat}

When the optical depth in the disk has dropped sufficiently, its heating becomes dominated by the direct irradiation of the host star. \cite{CG97} first studied the temperature structure of a passive disk in thermal equilibrium with the irradiation of its host star. Their results were then used by \cite{Cham09} to compute the density structure of his disk in the radiatively heated regime. The transition between these two regimes is referred to as the heat transition and is defined as the point where the temperatures derived by both heating sources is equal (see \cite{Cham09}, Equation 38). This transition point slowly moves inward as the disk ages and the accretion rate drops. As the accretion rate drops, the viscous heating rate is reduced and so the viscously heated region of the disk cools with time, until the disk is exclusively heated through direct irradiation. 

We have made a small modification to the formalism of \cite{Cham09} who assumed that the disk acted as a perfect blackbody in thermal equilibrium with the host star's radiation. As is well known, the perfect blackbody model does not accurately reproduce the SEDs of young stars, whereas a superheated upper atmosphere that heats the midplane \citep{CG97} more accurately reproduces SEDs. This more complicated model reduces the temperature at the midplane by a factor of $2^{1/4}$ when compared to the full blackbody temperature structure. 

From the superheated atmosphere model for radiatively heated disks \citep{CG97}, the midplane temperature is given by: \begin{align}
T= \left(\frac{\theta}{4}\right)^{1/4}\left(\frac{R_*}{r}\right)^{1/2} T_*,
\label{eqn15}
\end{align}
where $T_*$ and $R_*$ are the stellar temperature and radius respectively. The grazing angle $\theta$ of the light on the disk is dominated by the flaring of the disk, which has the form \citep{Cham09}: \begin{align}
\theta = r\frac{d}{dr}\left(\frac{H_{ph}}{r}\right),
\label{eqn16}
\end{align}
where $H_{ph}$ is the height of the disk where the gas optical depth from the host star reaches one. As in \cite{Cham09} and \cite{CG97} we assume that the disk is roughly vertically isothermal which results in a disk height of \citep{CG97}, \begin{align}
\frac{H_{ph}}{r} \simeq 4\left(\frac{T_*}{T_c}\right)^{4/7}\left(\frac{r}{R_*}\right)^{2/7},
\label{eqn17}
\end{align}
where $T_c$ is a constant with units of Kelvins (see \cite{Cham09}, his Equation 30). Combining eqs. \ref{eqn15}, \ref{eqn16}, \ref{eqn17} the resulting temperature profile is:\begin{align}
\left(\frac{T}{T_{rad}} \right) = \left(\frac{r}{s_0}\right)^{-3/7},
\label{eqn17x}
\end{align}
where \begin{align}
T_{rad} = \left(\frac{2}{7}\right)^{1/4} \left(\frac{T_*}{T_c}\right)^{1/7}\left(\frac{R_*}{s_0}\right)^{3/7} T_*.
\end{align}

An important distinction between the viscously heated regime and the radiatively heated regime is the lack of $\mdot$ dependence in the latter.  In the viscous regime the temperature of the disk cools as it ages (mass accretion rate decreases), however the radiative regime is passive. and the temperature does not change with time. As the mass accretion rate drops and the viscous regime cools the heat transition point moves inwards as radiative heating becomes more dominant. The density profile in the radiative regime has the form: \begin{align}
\Sigma = \Sigma_{rad}\left(\frac{\mdot}{\mdot_0}\right)\left(\frac{r}{s_0}\right)^{-15/14},
\label{eqn18}
\end{align}
where $\Sigma_{rad}$ is the surface density of the disk at the edge of the radiative region. We assume that the host star does not evolve in time as the disk accretes. This is a fair assumption for G-type stars but would need to be taken into account for M-type stars which spend up to an order of magnitude longer in their pre-main sequence stage of evolution \citep{Hay61}. The evolution of the disk mass due to viscous accretion is modified when a radiative region is present. Combining eqs. \ref{eqn10}, \ref{eqn18}, \ref{eqn12} and \ref{eqn13} we find: \begin{align}
\frac{dM}{dt} \equiv -\mdot = -\mdot_1\left(\frac{M}{M_1}\right)^{20/7},
\label{eqn19}
\end{align}
where $\mdot_1$ is the mass accretion rate when the radiative regime appears and $M_1$ is the disk mass at that time.

For the purpose of computing the temperature and density profiles at a given time we use the temperature profile (Equation \ref{eqn11x} or \ref{eqn17x}) that gives the higher temperature for a given radius, and the corresponding density profile (Equation \ref{eqn11} or \ref{eqn18}). The mass accretion rate in Equation \ref{eqn11} is computed by substituting the mass accretion rate solution of Equation \ref{eqn14} if no radiatively heated regime exists. Otherwise \ref{eqn19} is substituted into Equation \ref{eq:mdot02} and is used to evolve the mass accretion rate. The disk model is initialized at a time of 0.1 Myr with initial disk mass M$_0$ and size s$_0$. For further details see \cite{Cham09} or the Appendix section \ref{sec:calcmod}.
		% Disk model exist in this tex file

\begin{table}
\begin{center}
\caption{ Fiducial disk model parameters }
\begin{tabular}{c c}\hline
$M_{disk}(t=0)$ & $0.1$ M$_\odot$ \\\hline
$s_{disk}(t=0)$ & $66$ AU \\\hline
$M_{star}$ & $1.0$ M$_\odot$ \\\hline
$R_{star}$ & $3.0$ R$_\odot$ \\\hline
$T_{star}$ & $4200$ K \\\hline
$\alpha$ & $0.001$ \\\hline
$t_{in}$ & $0.1$ Myr \\\hline
$t_{dep}$ & $1$ Myr \\\hline
$t_{life}$ & $4.10$ Myr \\\hline
$L_{xray}$ & $10^{30}$ erg \\\hline
\end{tabular}\\
\label{tab:result01}
Note: $t_{in}$, $t_{dep}$ and $t_{life}$ are from Equation \ref{eq:mdot02}. $L_{xray}$ is the normalizing luminosity of the x-ray source in the system.
\end{center}
\end{table}

Table \ref{tab:result01} shows the fiducial values for our disk model parameters, including the luminosity normalization used in the radiative transfer of X-rays.

\section{ Disk Chemistry }\label{sec:Chem}

The chemistry in the disk is computed with a non-equilibrium, coupled, differential equation solver as seen in \cite{Berg03}, \cite{Fog11} and \cite{Cle14} and is based on the gas-grain chemical code: ALCHEMIC \citep{Sem10}. Generally, the abundance of a given chemical species $n(i)$ will evolve as \citep{Fog11}:\begin{align}
\frac{d n(i)}{dt} &= \sum_j\sum_l k_{jl}n(j)n(l) - n(i)\sum_j k_{ij} n(j),
\end{align}
which is to say that the abundance is determined by the sum of all producing reactions (of reactants $n(j)$ and $n(l)$) subtracted by the sum of all destructive reactions. The set of reactions (and their associative reaction rates $k_{ij}$) are taken from the Ohio State University Astrophysical Chemistry Group gas-phase model of 2008 \citep{Sea04} and was expanded on by adding unshielded photodissociation reactions from the UMIST 2006 database \citep{Wood07} by \cite{Fog11} and has since been updated with the photodissociation rates of the UMIST 2012 database \citep{McE03}. The chemical code and database have been well tested by these previous authors, and we have made no modifications to the core of the chemical code. The complete chemical database includes reaction rates for gas phase chemistry, grain surface reactions of hydrogen and water, photo-dissociation, X-ray and cosmic ray ionization, and grain absorption/desorption including thermal, cosmic ray and photodesorption. The chemical network includes 
5909 chemical reactions and 9 chemical elements: H, He, C, N, O, S, Si, Mg, Fe (see Table \ref{tab:02}).

\ignore{
Absent from this chemical model is the chemical processing of the dust grains. Instead, the dust grains act as a catalyst for the gas chemistry and as a seed for species to freeze on. In principle, knowing the chemical composition of the solids would be important to understanding the physical structure of super Earths and mini Neptunes. The composition of the solid core is less important to the global structure of Jupiter sized planets because the core mass makes up only a small percentage of the total mass (M$_{\rm core}$/M$_{\rm Jupiter} \sim 1 - 3 \%$). Generally, Gibbs free energy minimization methods are used to compute the chemical composition of the dust grains because the condensing rates of silicates are very high, with a mean time of approximately one hour \citep{Pig11}. This timescale is much smaller than any physical timescale in the disk, so equilibrium methods are appropriate. In \cite{APC16a} we combine this Gibbs free energy method with the disk model from \S 2 to compute the chemical 
abundances of the solids necessary to understand the structure of super Earths.}

The most prominent of the above reactions are driven through interactions between the gas and the local radiation field. Reactions like ion driven gas chemistry, photo-dissociation and photodesorption are much faster than neutral gas reaction because of a lack of an activation barrier between reactants. To determine the UV and X-ray radiation field, previous works have relied on an in-house Monte Carlo radiative transfer code that included both opacities due to dust and gas (\cite{WD01} and \cite{BB11} respectively) as well as line transfer for Lyman-alpha photons (see \cite{Berg03}), which can carry the majority of the UV flux through the disk.  We found that this algorithm was computationally expensive, as it requires an iterative calculation to determine the number density of molecular hydrogen and the resulting radiation field \citep{BB11x}. A much more efficient radiative transfer code was needed to compute the chemical profiles in many disk snapshots ($\sim 300$) in order to achieve the necessary 
temporal resolution discussed below.

Towards this end, we employ RADMC3D \citep{RADMC} to compute the radiation field, maintaining the same opacities as in the above sources. However the tradeoff is that we are neglecting the line transfer of Lyman-alpha photons, which means we are likely underestimating the total UV flux through our disk and the final abundance of some molecules that are dependent on UV photolysis to drive their production. For example the production of CN from the photolysis of HCN will obviously be reduced. This limitation is an issue when comparing computed disk astrochemistries to observations, a field that has seen major developments over the last decade. However, the molecules that have been detected or inferred in exoplanetary atmospheres are not as sensitive to the UV field. Since our focus is to compare our work with observations of abundances in exoplanetary atmospheres, neglecting Lyman-alpha photons will not significantly affect our results. In future works, we will improve both our disk model and radiative 
transfer schemes to more accurately replicate the chemical structure in accretion disks.

At every time step we initialize the chemical code with the temperature, surface density, and radiation field profiles calling the combinations of these profiles a `snapshot' of the disk. The chemical code uses a 1+1D method where the chemistry is computed at each radii individually, up to ten scale heights. In this method no chemical information is exchanged between radii, so effects like radial mixing are ignored. To minimize possible errors from radial mixing we resolve the temporal evolution of the chemistry to timesteps that are less than the viscous time in the disk at $1$ AU. This results in a timestep of approximately $13000$ years. With this choice of timestep, changes in the chemical structure between temporal snapshots of the chemical disk are due to changing disk properties like temperature and surface density rather than radial transport of material. 

As an additional approximation, we assume that the initial chemical abundances are the same for all disk snapshots, as shown in table \ref{tab:02}. In doing this, the chemical structure of any given snapshot does not depend on the chemical abundances computed in the previous snapshot. This assumption does have implications on the chemical structure that is computed, however the majority of the effect is seen in regions of the disk where photochemistry has the highest impact.  At the midplane, differences between a snapshot that is initialized with the values of table \ref{tab:02} and the same snapshot that is initialized with the abundance from the previous one are less than $\sim 10\%$ in the region of the disk where atmospheric gas accretion takes place (at radii within $1$ AU). This assumption reduces our average computation time by a factor of 20 because successive snapshots can be run in parallel on a supercomputing cluster. This decrease in 
computing time is necessary for the time resolution we required for this work and will become increasingly important when a parameter study of our model is done. The initial concentration of the chemical species are shown in table \ref{tab:02} and was taken from \cite{Fog11} and \cite{AH99}.

\begin{table}
\centering
\caption{Initial abundances relative to the number of H atoms. Included is the initial ratio of carbon atoms to oxygen atoms (C/O) and the initial ratio of carbon atmos to nitrogen (C/N).}\label{tab:02}
\begin{tabular}{l l l l}\hline\hline
Species & Abundance & Species & Abundance \\\hline
H$_2$ & $0.5$ & H$_2$O & $2.5\times 10^{-4}$ \\
He & $0.14$ & N & $2.25\times 10^{-5}$ \\
CN & $6.0\times 10^{-8}$ & H$_3^+$ & $1.0\times 10^{-8}$ \\
CS & $4.0\times 10^{-9}$ & SO & $5.0\times 10^{-9}$ \\
Si$^+$ & $1.0\times 10^{-11}$ & S$^+$ & $1.0\times 10^{-11}$ \\
Mg$^+$ & $1.0\times 10^{-11}$ & Fe$^+$ & $1.0\times 10^{-11}$ \\
C$^+$ & $1.0\times 10^{-9}$ & GRAIN & $6.0\times 10^{-12}$ \\
CO & $1.0\times 10^{-4}$ & N$_2$ & $1.0\times 10^{-6}$ \\
C & $7.0\times 10^{-7}$ & NH$_3$ & $8.0\times 10^{-8}$ \\
HCN & $2.0\times 10^{-8}$ & HCO$^+$ & $9.0\times 10^{-9}$ \\
C$_2$H & $ 8.0\times 10^{-9}$ & C/O & $0.288$ \\
& & C/N & $4.09$ \\
\hline
\end{tabular}
\end{table} 

\begin{figure*}
\centering
\includegraphics[width=\textwidth]{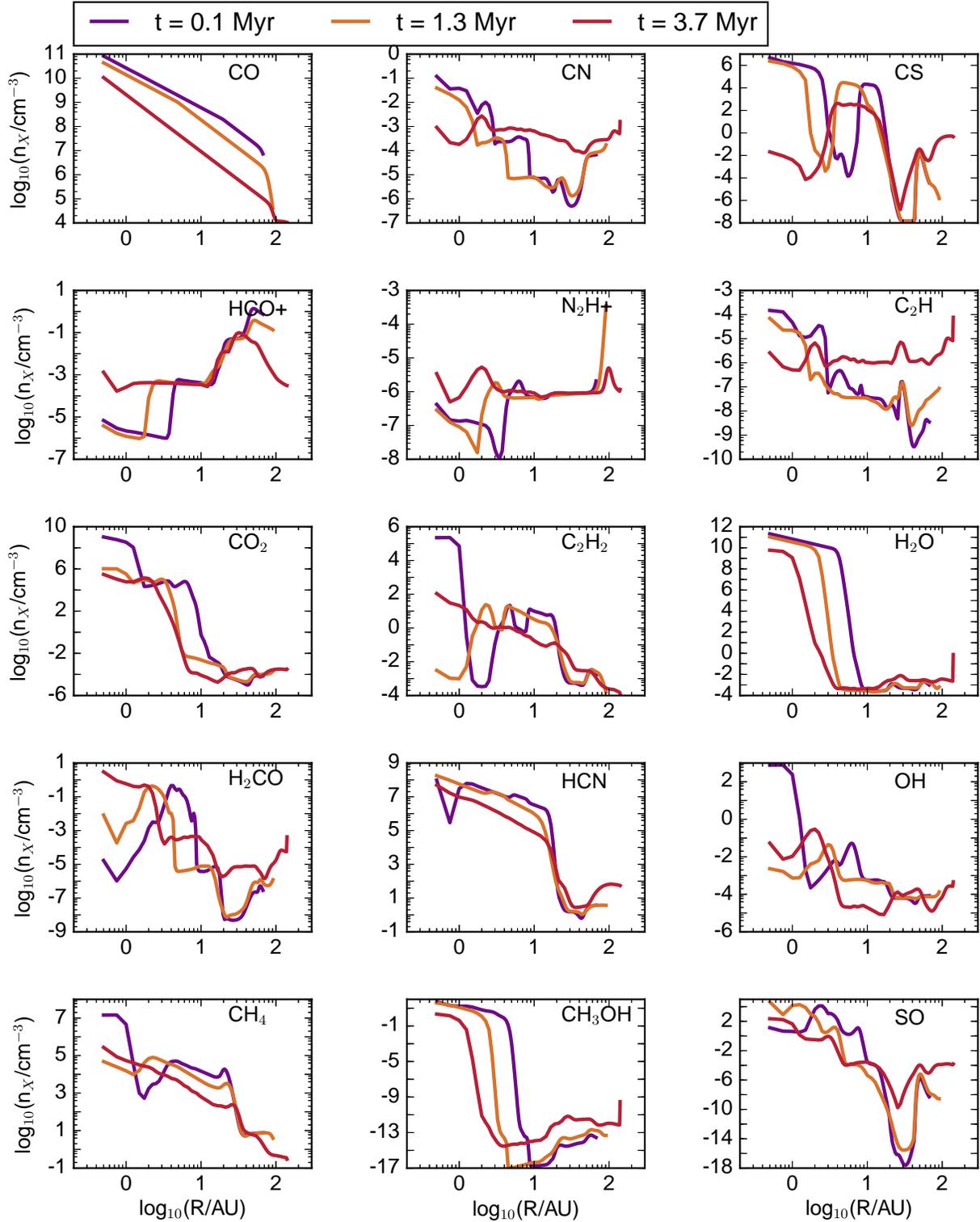}
\caption{Number density at the midplane of the disk as a function of radius for a number of important gas molecules in our chemical model.}
\label{fig:gaschem}
\end{figure*}

\begin{figure*}
\centering
\includegraphics[width=\textwidth]{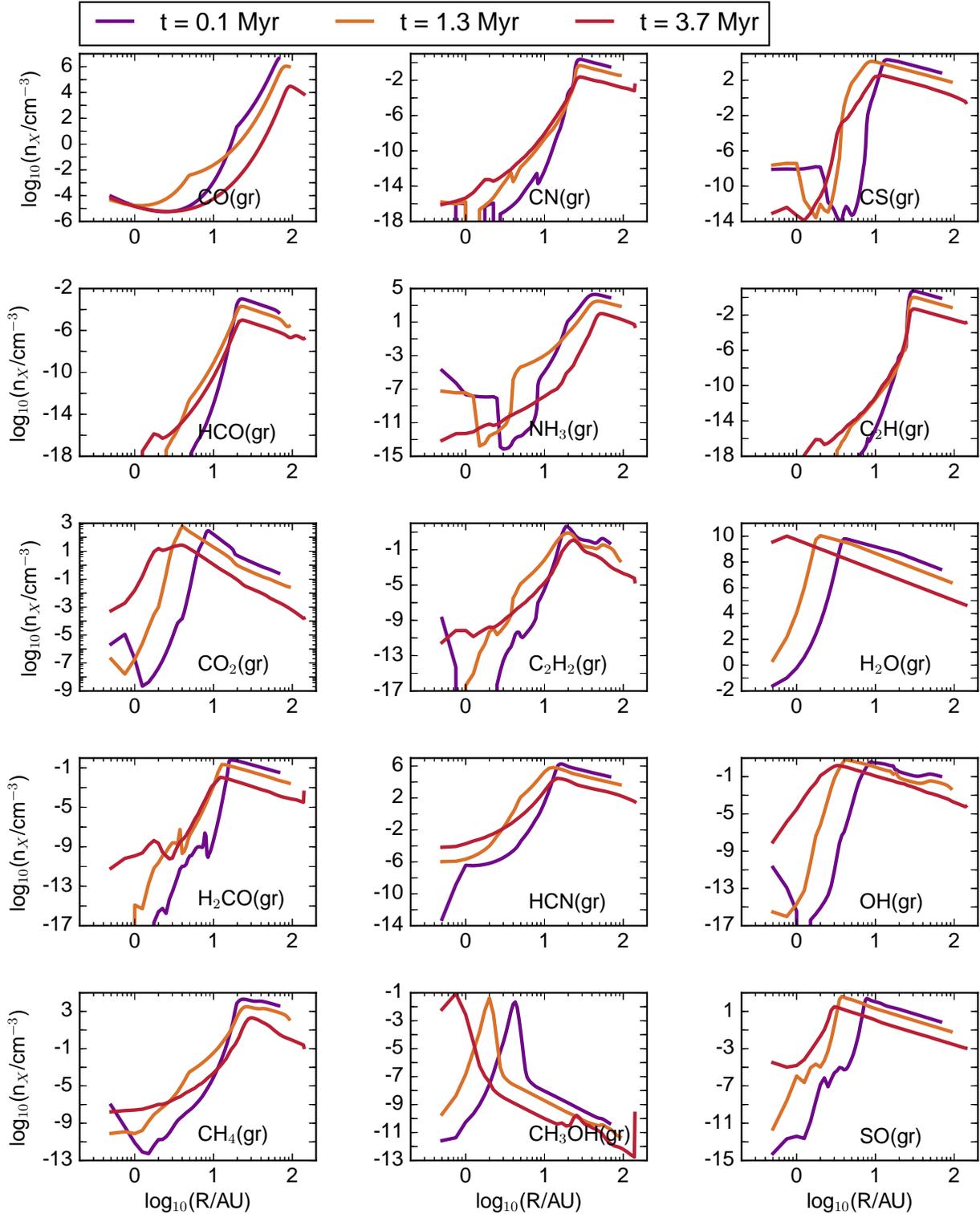}
\caption{Same as in Figure \ref{fig:gaschem} except for molecules frozen on grains. The only different chemical species is in the middle of the second row, as no frozen component of N2H+ exist in the model. Instead, the frozen component of NH3 is shown.}
\label{fig:icechem}
\end{figure*}

Figures \ref{fig:gaschem} and \ref{fig:icechem} show the time evolution of the radial profiles of a set of gas and ice species in the chemical code. Generally we see strong variations in the abundance of most molecules over all radii as the disk ages. At low radii, where the disk is viscously heated the decreasing temperature causes features in molecular abundances to shift inward while maintaining most of its shape. Examples of this type of evolution is in both the gas and ice species of H$_2$O, CH$_3$OH, CO and CO$_2$, as well as the gas species of HCO+ and N$_2$H+. On the other hand at larger radii the temperature structure becomes passive in the radiatively heated regime of the disk and the evolution of the surface density becomes the primary driver of evolution. In the radiative regime, surface density reduction caused by the accretion of matter through the disk results in a larger flux of high energy photons, leading to variations in molecular abundance due to radiation driven chemistry. For example, 
the rate of reaction of photodissociation reactions has the form \citep{Fog11} \begin{align}
k_{photodiss} &= \int \frac{4\pi \lambda}{hc} \sigma(\lambda) J_\lambda(r,z)d\lambda ~s^{-1}.
\end{align}
As the surface density decreases the flux of photodissociating photons $J_\lambda(r,z)$ increases, leading to stronger irradiation of the midplane at all radii and a higher rate of gas species destruction. 

This higher irradiation is also tied to the overall destruction of some species like CH$_3$OH at higher radii. While hydrogen and water are allowed to react on grains, other species are protected from further chemical evolution when they freeze onto dust grains because there are no other surface reactions included in the model \citep{Fog11}. However at higher radii species become more susceptible to desorption caused by high energy photons or cosmic rays. Once in the gas phase, these species are destroyed through reactions with ions or dissociation caused by cosmic rays. The rate of photodesorption depends both on the local radiation field as well as the properties of the dust species that are present. 

\ignore{
It has the form \citep{Fog11},\begin{align}
k_{photodesorp} &= F_{UV} Y \frac{\sigma_{gr}}{N_{sites}} N_p\frac{n(i)}{n_{ice}}N_m^{-1} ~s^{-1},
\end{align}
increases with increasing flux $F_{UV}$ of photons. The reaction rate also depends on the yield $Y$ of a photodesorption event, the number of sites on the grain ($N_{sites}$) and $\sigma_{gr}$ is the cross sectional area of the grain. 

This last dependence becomes important when considering the effect that grain growth and settling will have on the chemistry in the disk. Numerical simulations have shown that the smallest grains are quickly depleted through coagulation \citep{DD05,Gara13} and the resulting average grain size in the disk tends to grow in time. In future work, we will investigate the impact that these dust coagulation models have on the global chemical structure of the disk. The number of monolayers $N_m$ and the correction factor $N_p=2$ are determined through laboratory experiments by \cite{O07}. The ratio between the gas abundance of the species $n(i)$ and the total ice abundance $n_{ice}$ suggest that (as expected) the more species in the ice phase, the more likely a photodesorption event will happen.}

In particular, this process appears important for methanol (CH$_3$OH, row 5, column 2 in Figures \ref{fig:gaschem} and \ref{fig:icechem}). From the Figures, we observe a freeze out at disk radii of $\sim$ 5, 2.5 and 1 AU at 0.1, 1.3 and 3.7 Myr respectively, tracing the cooling of the disk as observed in other plots of other volatiles like water. At larger radii in each of the snapshots of the disk, there is a strong depletion of the frozen component of CH$_3$OH without an increase in the gas abundance at the same location. The survivability of methanol has interesting impacts on observations and is important for the formation of biologically relevant molecules like glycolaldehyde \citep{J12}.

Water undergoes a simpler time evolution and is in a class of species that is less sensitive to the presence of high energy radiation. Water vapour undergoes a strong reduction as seen in the ninth (row 3, column 3) panel of Figure \ref{fig:gaschem} which is correlated with the strong increase of water ice in the ninth panel of Figure \ref{fig:icechem}. This process is dominated by the equilibrium of freeze out and sublimation of water. The half height of this strong reduction will be where we define the location of the Ice line trap (see below).

\section{ Planet Formation and Migration }\label{sec:Plnt}

When a sufficiently small planetary core ($\leq$ a few Earth masses) is embedded in the disk, the disk structure is not drastically changed. However, spirals waves are excited through constructive interference of density waves at Lindblad resonances \citep{W97}. These density waves exert torques on the planet which generally result in a loss of angular momentum from the planet to the disk, forcing the planet to migrate inwards (this is known as Type-I migration). The direction of the migration is set by both the surface density and temperature profiles. For simple disk distributions, only the two closest Lindblad resonance positions contribute to the net torque on the planet. One resonant location is at a larger disk radius than the planet, while the other is at a smaller radius (outer and inner resonances respectively).  The outer resonance lies closer to the planet than the inner resonance. It has a stronger effect transferring angular momentum from the planet to the disk, causing net inward migration. 
Early on in planet formation theory, it was realized that the timescale for planet formation ($\sim 10^6$) was much longer than the timescale for Type-I migration ($\sim 10^5$, \cite{W97}, \cite{M06}). Current three dimensional migration calculations reproduce these low migration timescales, and still need to be reduced by a factor of between 10-100 \citep{M06} to account for the survival of planets. 

There is a second source of torque on the migrating planet which contributes to Type-I migration: the corotation torque. Also known as the horseshoe drag, corotation torques are caused by gravitational interactions between particles on horseshoe orbits around the planet \citep{W91}. Horseshoe orbits occur for fluid packets that have orbital frequencies similar to that of the planet. Since a fluid packet is initially at a larger orbital radius than the planet, as the planet catches up to the packet it is perturbed to a lower orbit and loses angular momentum. The packet at the lower orbit now moves faster around its orbit than the planet, catching up to the planet within an orbital period. At this point the planet perturbs the packet again up to a higher orbital radius, regaining the lost angular momentum from the first encounter of the horseshoe orbit. 

Since a disk is generally cooler the further it is from the star, a fluid packet that moves from the outer region of the horseshoe orbit to the inner region will be cooler than the surrounding gas. If the fluid packet returns to the opposite side of the horseshoe orbit without changing (ie. no turbulent mixing) there is no net angular momentum transfer between the fluid packet and the planet over the full horseshoe orbit described above. Under such a situation the corotation torque is said to be `saturated'. This occurs if the viscous diffusion timescale are longer than the libration time of the horseshoe orbit. In the opposite case the torque is called `unsaturated'. The corotation torque, when unsaturated, can overcome the Lindblad torques in most disk set-ups, leading to slowed or outward migration. To determine if the corotation torque is saturated, one must compare the libration and the viscous timescales. The libration time and viscous time has the form \citep{MasPap03}:
\begin{align}
t_{Lib} &= \frac{4\pi r_p}{1.5\Omega_p x_s} \label{eq:tlib} \\
t_{vis} &= \frac{x_s^2}{3\nu} \label{eq:tvis},
\end{align}
respectively. Here $r_p$ is the radial position of the planet and $\Omega_p$ is the Kepler frequency at $r_p$. The viscosity $\nu$ is defined in Equation \ref{eq:in01b}. The distance traversed by the fluid packet through the horseshoe orbit $x_s$ is defined by\begin{align}
x_s &= 0.96 r_p \sqrt{\frac{M_p r_p}{M_* h_p}}
\end{align}
where the $M_p$ is the mass of the planet and $h_p$ is the scale height of the disk at the location of the planet. 

\begin{figure}
	\centering
	\includegraphics[width=0.5\textwidth]{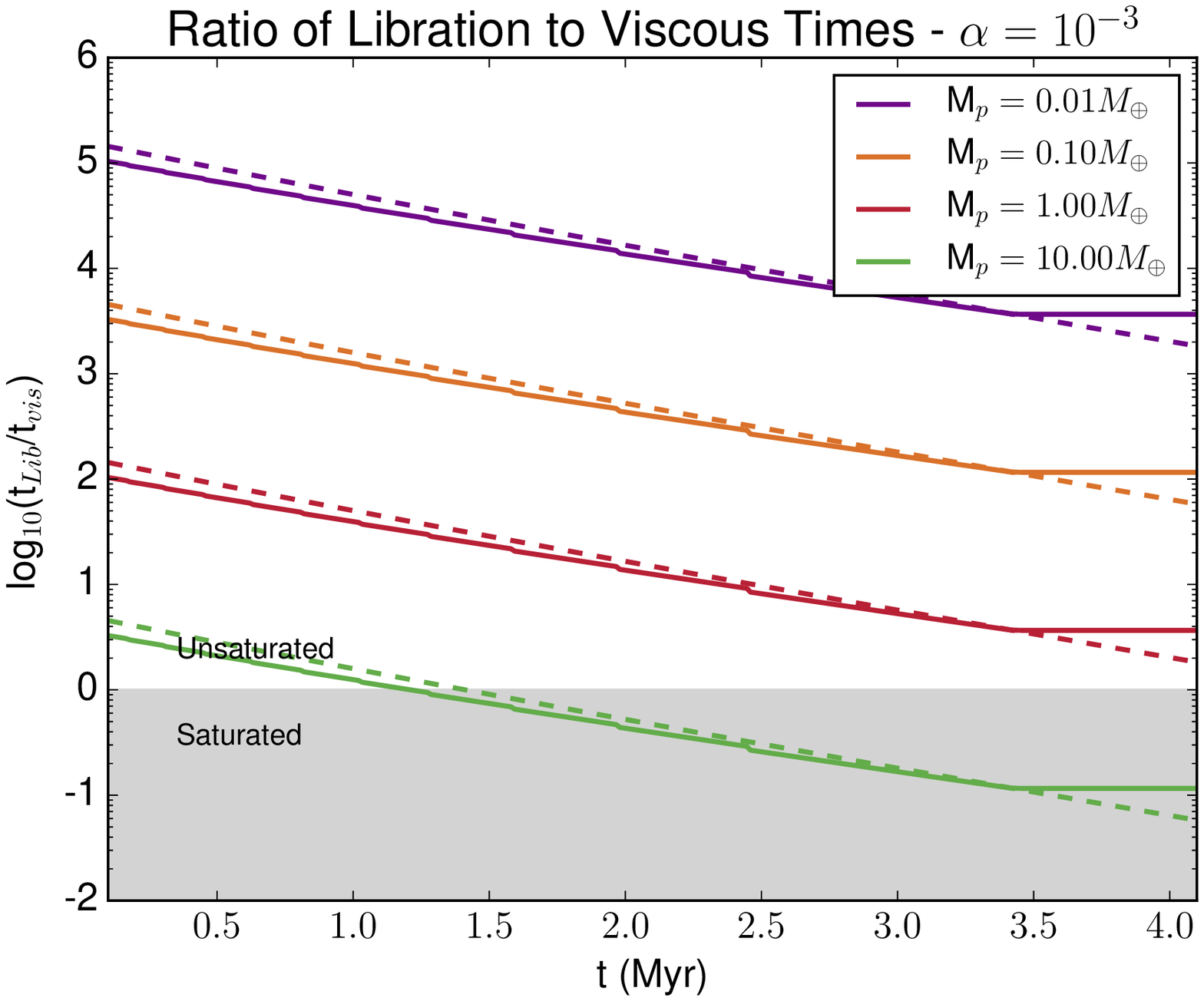}\\
	\includegraphics[width=0.5\textwidth]{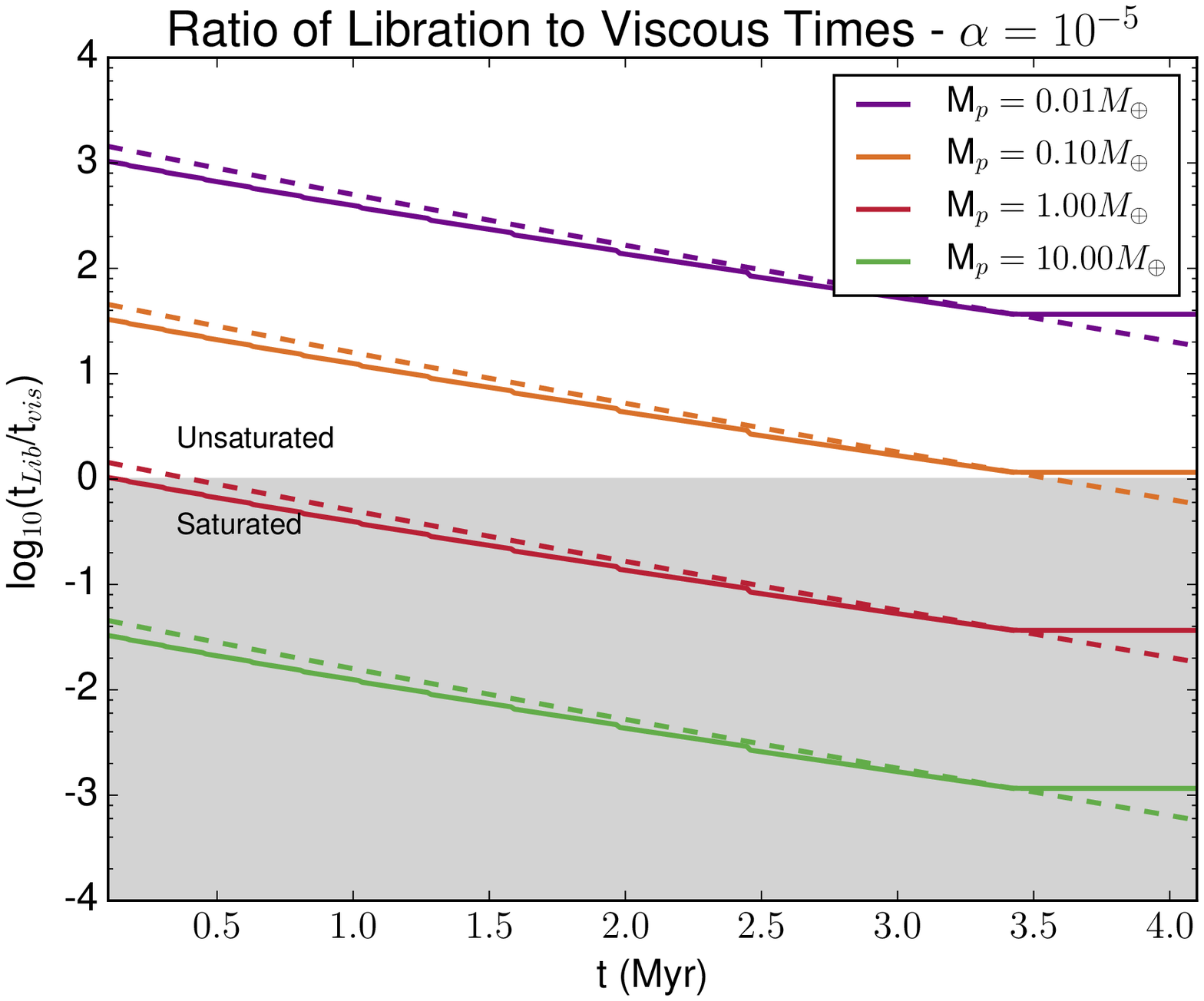}
	\caption{Ratio of libration time to viscous time for an active ($\alpha_{turb} = 10^{-3}$) and dead ($\alpha_{turb} = 10^{-5}$) disk. In both cases, the location of the planet is determined by the location of the ice line (solid line) and heat transition traps (dashed line) for a fiducial disk model. The gray region denotes the parameter space where the corotation torque is saturated.}
	\label{fig01b}
\end{figure}

In Figure \ref{fig01b} the ratio of these timescales are shown for four different planet masses for our evolving fiducial disk. The smallest bodies are trapped over a long time by an unsaturated corotation torque for both turbulently dead and turbulently active disk (denoted by an $\alpha_{turb}$ of $10^{-5}$ and $10^{-3}$ respectively). If the libration time is larger than the viscous time the turbulence is sufficiently strong to mix the packet of fluid with the surrounding fluid, keeping the corotation torque unsaturated. In the turbulently active disk, only the most massive planets ($\gtrsim 10$ M$_\oplus$) will see a saturated corotation torque at times $\gtrsim 1$ Myr; however at this higher mass Type-II migration will set in and this test does not apply. In the turbulently dead zone a $1$ M$_\oplus$ planet will be trapped by the corotation torque for a few hundred thousand years before the corotation torque saturates. In general a protoplanet will begin its evolution on the trajectory of the smallest 
planet in Figure \ref{fig01b}. 
As a planet 
accretes matter it will shift from one line to another, entering into a saturated region earlier than a planet that is not undergoing mass accretion. 

Strong density, temperature or vortensity gradients can reverse the direction of Lindblad torques or can strengthen the impact of corotation torques, halting Type-I migration \citep{P11}. We call these inhomogeneities `planet traps' and we select discrete disk properties to define their position and evolution. A complete picture of the evolution of our three planet traps of interest (see below) has been computed by \cite{HP11}. The three planet traps of interest in our model are the water ice line, heat transition, and dead zone. In what follows, we extend the work of \cite{HP11} by including the above mass dependent saturation test when the ice line and heat transition traps are located within the dead zone of the disk.

\subsection{ Defining a Planet Trap }

Our disk chemistry calculations allow us to improve on previous methods of defining the location of the above planet traps, as two of the traps can be defined by the chemical structure of our protoplanetary disk model.

\subsubsection{ Heat Transition Trap }\label{sec:HTrap}

\begin{figure}
\begin{center}
\begin{overpic}[width=0.5\textwidth]{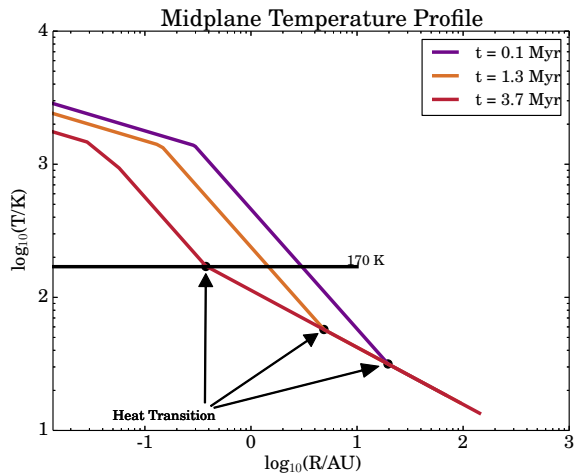}
\end{overpic}
\caption{The temperature profile (Equations \ref{eqn11x} and \ref{eqn17x} along with the modification from Equation \ref{eq:ht01} on Equation \ref{eqn11x}) for three snapshots are shown for the fiducial disk (Table \ref{tab:result01}). The black dot and arrows denote the defined location of the heat transition trap for three temperature profiles shown here. The intersection between the horizontal line and the temperature profile at $170K$ denotes an approximate location of the water ice line in this model. In the radiative region, the temperature profile becomes independent of time, and the profiles overlap in this region.}
\label{fig02}
\end{center}
\end{figure}

The simplest trap to identify is the heat transition trap, which is located at the transition radius between a region primarily heated through viscous heating and a region heated through direct irradiation. This radius is defined in \cite{Cham09} and is identified for three computed temperature profiles in Figure \ref{fig02}.

In Figure \ref{fig02} we show the evolution of the midplane temperature profile for our fiducial disk model at $0.1, 1.3$ and $3.7$ Myr. The times are selected to represent the disk at its `early', `middle' and `late' ages for the fiducial model presented. The lowest temperatures are due to direct irradiation and is time independent, as we assume that the star is not evolving during the full lifetime of the disk, and that the disk is isothermal from the disk photosphere down to the midplane. At the highest temperatures its functional form is modified by what is known as the evaporative region in \cite{Cham09}. In this region the opacity is allowed to vary with temperature while the disk is heat through viscous evolution. Aside from this innermost regime, both the viscous and irradiated regimes are assumed to have the same opacity. As in \cite{Cham09} we note that a constant opacity is not a realistic representation for an astrophysical accretion disk, and indeed a more complete representation of opacity has 
been carried out by \cite{St98}, \cite{Bea97} \& \cite{BL94}. However this picture is further complicated by dust settling and growth and a full description evades an analytical solution, as in the combination of Equations \ref{eqn11x}, \ref{eqn17x}, \ref{eqn11} and \ref{eqn18}.  In other astrochemical works, dust settling has been completely ignored \citep{Wal14} or has been been computed as a consequence of the equilibrium between gravitation and turbulent mixing \citep{Cle14,Hel14} in a passive disk. A time dependent treatment of dust evolution, which is susceptible to processes like radial drift \citep{Birn12}, has yet to be connected to chemical evolution. We leave the analysis of the impact of evolving dust structure to a future paper. 

In the context of our analytic disk model, we choose to keep the dust opacity constant at a temperature below the evaporation temperature of dust ($\sim 1380$ K). The full functional form of the opacity is \citep{Cham09}, \begin{align}
\kappa &= \kappa_0 & T < T_e\nonumber\\
\kappa &= \kappa_0\left(\frac{T}{T_e}\right)^n\quad & T > T_e,
\label{eq:ht01}
\end{align}
where $T_e = 1380$ K is the evaporation temperature and $n = -14$ \citep{St98}. This equation is combined with Equations \ref{eqn04} and \ref{eqn05} where it replaces the constant $\kappa_0$. With this modification, the temperature profile changes from the standard profile in the viscously heated region of the disk at small radii.

The location of the heat transition trap is highlighted three times in Figure \ref{fig02}. As time goes on the position of the heat transition moves inward as the viscous region cools and more of the disk is heated by direct irradiation. The origin of this trap is generally due to an entropy gradient across the trap, on which the strength of the corotation torque depends. 

\cite{P11} showed that the total torque from both the Lindblad resonances and unsaturated corotation torques are, \begin{align}
\frac{\gamma\Gamma}{\Gamma_0} &= -2.5 - 1.7 \beta + 0.1 s + 1.1\left(\frac{3}{2} - s\right) + 7.9\frac{\xi}{\gamma},
\label{meth01}
\end{align}
where $s$ and $\beta$ are the exponents in Equation \ref{eqn01}, $\xi \equiv \beta - (\gamma-1)s$ is the power law exponent for the entropy and $\gamma$ is the ratio of specific heats. $\Gamma_0 = (q/h)^2\Sigma_pr^4_p\Omega_p^2$ is the standard normalization of torques on a planet at position $r_p$ of the planet. In the \cite{Cham09} model the relevant values in both heating regimes are shown in Table \ref{tab:01}.

Substituting power law indices in both the viscous and irradiated regimes shows that the sign of the total torque is positive in the viscous region, so that the planet will migrate outwards. Meanwhile, in the radiative region the sign of the total torque is negative and the planet will migrate inward. The reversal of the total torques cause the planet to be trapped at the transition point between the two heating sources. 
%\begin{wraptable}{r}{0.26\textwidth}
\begin{table}
\centering
\caption{Power law exponents and ratio of specific heats used in the Chambers model. The last row denotes the sign of the total torques on the planet locally around the heat transition.}
\begin{tabular}{|l|c|c|} \hline
& Viscous & Radiative \\\hline
s & $\frac{3}{5}$ & $\frac{15}{14}$ \\\hline
$\beta$ & $\frac{9}{10}$ & $\frac{3}{7}$ \\\hline
$\gamma$ & $1.4$ & $1.4$ \\\hline
$\xi$ & $\frac{33}{50}$ & $0$ \\\hline
$sign(\Gamma)$ & $+1$ & $-1$ \\\hline
\end{tabular}
\label{tab:01}
%\end{wraptable}
\end{table}
As seen in Figure \ref{fig02} the heat transition begins out beyond $10$ AU and moves inward of $1$ AU by $2-3$ Myr. A trapped planet will travel over an order of magnitude in radius through the disk, and should sample a wide range of gas with varying chemical abundances.

\subsubsection{Ice Line Trap}\label{sec:ITrap}

In the previous work of \cite{HP11} the ice line trap was defined as the position in the disk that has a temperature equal to the sublimation temperature of water vapour (directly from ice). This sublimation temperature is approximately $170$ K and is denoted by the horizontal line in Figure \ref{fig02}. The traditional methods assumed that the only limiting effect on the amount of water vapour was the local temperature of gas, and the relative rates of thermal absorption and desorption. To improve on this previous method we use the results of our astrochemical code to locate the radial position of the ice line at the midplane.

\begin{figure}
\begin{center}
\begin{overpic}[width=0.5\textwidth]{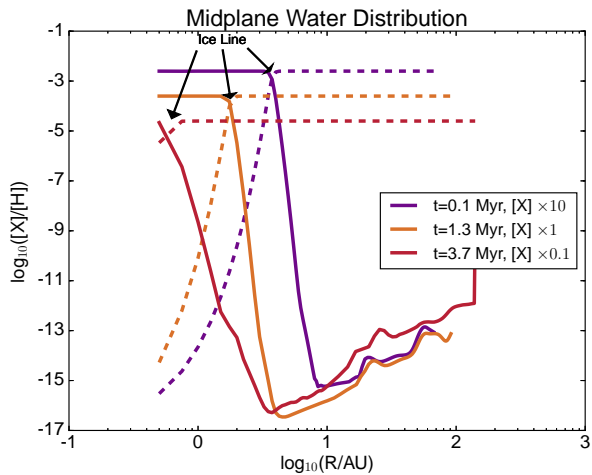}
\end{overpic}
\caption{Number density of water vapour (solid line) and water ice (dashed line) relative to hydrogen - offset to minimize overlap. In reality each curve has the same maximal abundance. The ice line refers to the transition point between the disk's water content being primarily in vapour form to primarily in ice. As the disk ages and cools, the ice line moves inward on the viscous timescale, eventually disappearing beyond our resolution range.}
\label{fig03}
\end{center}
\end{figure}

In Figure \ref{fig03} we show the abundance of water vapour (solid line) and water ice (dotted line) relative to the abundance of hydrogen at the same times shown in Figure \ref{fig02} for our fiducial disk. At the earliest times the abundance of water vapour, relative to hydrogen, does not evolve other than across the ice line. So in the figure, we have multiplied each line by a factor of 10, 1, and 0.1 for the times $0.1$, $1.3$ and $3.7$ Myr respectively to better display each curve. The water ice line is defined by the position in the disk where the abundance of water vapour and ice is equal. By using a full chemical code, we gain an immediate insight into other effects that might play a role in determining the position of the ice line outside of simply the gas temperature, such as pressure and radiative effects.

At the ice line, the opacity $\kappa$ is reduced as the dust grains are coated with water ice, and the associated cooling rate of the gas is increased (since $D\sim 1/\kappa$, \cite{B13}). Coupled to the cooling rates, the local temperature and thus the disk scale height, is reduced ($H \sim c_s\sim T^{1/2}$). In the standard $\alpha$-disk model the mass accretion through the disk ($\mdot \sim \nu \Sigma$) is assumed to be constant. Because the viscosity is dependent on the local gas temperature ($\nu \sim c_sH\sim T$), a reduced temperature results in an enhancement in the local surface density at the ice line to maintain a constant mass accretion. Both a modified temperature structure and density enhancement impact the direction of the net torques by locally increasing the scale height of the disk at radii just larger than the trap location. This increased scale height results in a large negative temperature gradient ($\beta > 0$) and a large positive surface density gradient ($s < 0$) across the ice line 
trap, and a strong negative torque outside of the trap (see Equation \ref{meth01}).

Figure \ref{fig02} also shows that two traps might cross in space at some point in the disk history, which appears to occur at about $3.7$ Myr in the fiducial model. Two interesting dynamic questions can be asked: what happens to two planets that are forming simultaneously in the two traps when they cross? If only one planet is forming in one of the traps, does the planet `jump' to the other trap when they cross? Both of these questions are beyond the scope of this work but offer an interesting future addition to our model, where forming planets are allowed to interact dynamically with each other.

\subsubsection{ Dead Zone Trap }\label{sec:DTrap}

As was outlined in \S 2 we focus on disk turbulence that is driven by the magnetorotational instability (MRI). The ability of the MRI to excite turbulence is dependent on the magnetic field's coupling to the ionized gas in the accretion disk. This ability is often characterized by the Ohmic Elsasser number (Equation \ref{eq:in02}) which compares the rate of growth of MRI driven turbulence to the rate that energy is dissipated through Ohmic diffusion. As shown in Equation \ref{eq:in03} the Ohmic resistivity depends on the disk electron fraction which we compute by comparing the abundance of electrons to the abundance of hydrogen from the chemical code.

\begin{figure}
\begin{center}
\begin{overpic}[width=0.5\textwidth]{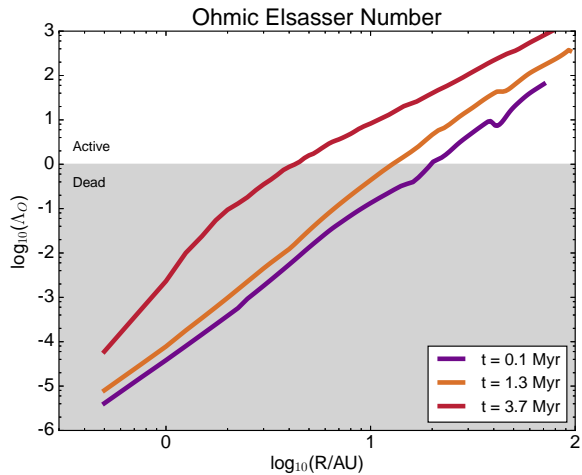}
\end{overpic}
\caption{Ohmic Elsasser number over three disk snapshots. The line between the white and grey regions denote the critical Ohmic Elsasser number $\Lambda_{O,crit}\equiv 1$ above which the disk would be turbulently active due to the MRI. The dead zone trap is denoted by the location where the Ohmic Elsasser number first (from the right) crosses into the dead region. }
\label{fig04}
\end{center}
\end{figure}

When the disk ionization is high enough that the Ohmic Elsasser number is above a critical value of $\Lambda_{O,crit} \equiv 1$ the disk is turbulently active from the MRI \citep{BH94,Simon13,Fro13}. We define the location of the dead zone trap as the radius in the disk where $\Lambda_O = 1$ and track this radius. In the dead zone the turbulent parameter ($\alpha_{turb}$) drops by about two orders of magnitude relative to the active region of the disk and the disk accretion rate due to turbulence is drastically decreased. In the dead zone the dust scale height is lower than in the active zone, thereby presenting a `dust wall' to the incident radiation. The back scattered radiation from the wall produces a radial temperature inversion and a thermal barrier to Type I migration \citep{HP10}.

In Figure \ref{fig04} we show the Ohmic Elsasser number in the fiducial model at the same times as in Figure \ref{fig02}. We also show the Ohmic Elsasser numbers which suggest where the disk would be active (white region) and where we expect it to be dead (gray region). As the surface density of the disk drops the flux of ionizing radiation can increase, enhancing the electron fraction and Ohmic Elsasser number. The electron fraction is computed by our astrochemical code and is dependent on the local radiation field, as well as the average size of the dust grains. In this fiducial model the dust grains are completely mixed with the gas and do not undergo any size evolution, remaining at a radius of $\sim 4~\mu$m for the lifetime of the disk. This size represents an average grain in the MRN distribution \citep{Math77} weighted by a freezing efficiency factor that is found in the chemical code. This factor affects the absorption rate of gas species onto grains and scales with the grains size. Larger grains 
have less surface area for absorption per mass of dust, and so the freezing efficiency is reduced.

As was outlined earlier, the likelihood that a planet will be trapped by the combined effect of the Lindblad and corotation torques depends on the magnitude of the $\alpha_{turb}$ parameter at a particular location of the disk. This implies that the location of the dead zone will also determine whether the corotation torque responsible for the other traps will be saturated or unsaturated. If either the ice line or heat transition traps exist within the dead zone of the disk (where we assume $\alpha_{turb} = 10^{-5}$) our analysis has shown that the corotation torque could saturate for a large enough planet. With the saturated corotation torque the semi-major axis of the young planet evolves only through Lindblad torques in a standard Type-I migration scenario. 

The magnitude of the total torque is \citep{Exo10},\begin{align}
\Gamma &= C_\Gamma \Sigma \Omega_p^2 a^4 \left(\frac{M_p}{M_s}\right)^2\left(\frac{a}{|\Delta r|}\right)^2.
\label{eq:torq01}
\end{align}
When the scale height of the disk at the planet location $H_p$ is less than the Hill radius $R_H$ of the planet the gravitational effect of the planet on the gas is larger than the internal pressure of the gas and the relevant length scale of the density gradient is $|\Delta r| = R_H$. The opposite is true when the effects of gas pressure are larger than gravity, so we write $|\Delta r| = max(H,R_H)$ to account for both limits. The constant $C_\Gamma$ has been computed through numerical simulations of Type-I migration, it has a value of $C_\Gamma = -(2.34 - 0.10 s)$ \citep{Exo10} for the saturated corotation torque.

The radial evolution of the planet is $\dot{r}_p/r_p = \Gamma/J_p$ where $J_p = M_p \Sigma_p r_p^2$ is the orbital angular momentum of the planet, leading to,\begin{align}
\frac{1}{r_p}\frac{dr_p}{dt} &= \frac{C_T \Sigma \Omega_p r_p^2}{M_p} \left(\frac{M_p}{M_s}\right)^2\left(\frac{r_p}{max(H,R_H)}\right)^2.
\label{eq:torq02}
\end{align}

If the ice line and heat transition trap are within the dead zone we calculate the ratio of the libration and viscous time to determine whether or not the planet would be trapped at each timestep. If the ratio of the libration time (Equation \ref{eq:tlib}) is larger than the viscous time (Equation \ref{eq:tvis}) the radial position of the forming planet is defined by the location of the trap.

This analysis is a recent addition to our model because of our chemically motivated method of inferring the location of the dead zone. Previous works relied on an analytic method of determining the ionization structure in the disk. It was generally found that the radial position of the dead zone was well within the other two traps for the entire disk lifetime \citep{HP13} or evolved inwardly of the two traps quickly enough \citep{APC16a} that such an analysis was not necessary.

Generally we assume that the radial location of the trapped planet will follow the location of its planet trap exactly. At every timestep we check the ratio of the libration time to viscous time, and in the case that the ratio $t_{Lib}/t_{vis}$ is less than one the radial evolution is defined by Equation \ref{eq:torq02}.

\subsection{ Planet Formation Model }

Our planet formation model is based on the standard core accretion model of \cite{IL04}. The planet grows by first growing a solid core through oligarchic growth \citep{KI02} followed by gas accretion onto the established core regulated by the Kelvin-Helmholtz (KH) timescale \citep{I00}. The core begins at one hundredth the mass of Earth and quickly migrates into one of the three planet traps. The core then grows at the radii defined by the planet trap until it reaches a mass large enough to open a gap and begin Type-II migration.

In the core accretion model the core undergoes three phases of accretion. The first phase is a rapid solid oligarchic growth, with growth timescales of $O(10^5)$ yr. The mass accretion rate of the solid core is defined as (\cite{IL04}, their equations 5 and 6): \begin{align}	
\frac{dM_c}{dt} &= \frac{M_c}{\tau_{c,acc}} \nonumber\\
	&\simeq \frac{M_c}{1.2\times 10^5} \left(\frac{\Sigma_d}{10 {\rm gcm}^{-2}}\right)\left(\frac{a}{1 {\rm AU}}\right)^{-1/2} \left(\frac{M_c}{M_\oplus}\right)^{-1/3}\left(\frac{M_s}{M_\oplus}\right)^{1/6}\nonumber\\
	&\times \left[\left(\frac{\Sigma_g}{2.4\times 10^3 {\rm gcm}^{-2}}\right)^{-1/5}\left(\frac{a}{1 {\rm AU}}\right)^{1/20}\left(\frac{m}{10^{18} {\rm g}}\right)^{1/15}\right]^{-2} {\rm g~yr}^{-1},
\label{eq:acc01b}
\end{align}
where $\Sigma_g$ and $\Sigma_d$ is the surface density of gas and dust respectively, $M_c$ is the mass of the core and $m$ is the average mass of the accreting planetessimals. 

During oligarchic growth little to no gas is accreted since the accreting oligarchs keep the core too hot. When the core reaches its `isolation mass', defined by \cite{IL04} as the maximum available solid mass within $10$ Hill radii of the core, solid accretion stops and a phase of slow ($O(10^6)$ yr) gas accretion begins. The gas accretion is regulated by the KH timescale, parameterized by \citep{HP12}, \begin{align}
\tau_{KH} \simeq 10^c {\rm yr}\left(\frac{M_p}{M_{\oplus}}\right)^{-d},
\label{eq:acc01}
\end{align}
so that the gas accretion rate has the form, \begin{align}
\frac{dM_p}{dt} \simeq \frac{M_p}{\tau_{KH}},
\label{eq:acc01c}
\end{align}
where $M_p$ is the total mass of the forming planet, and the parameters $c=9$ and $d=3$ are fiducial values \citep{IL04} that adequately reproduce the mass-period relation for the population of observed exoplanets \citep{HP13}. If the forming planet accretes enough mass for the KH timescales to drop below about $10^5$ yr the planet enters into a phase of unstable gas accretion, where it rapidly draws down the remaining gas in its feeding zone. 

At some point before this final accretion phase the planet becomes massive enough to open a gap in the disk and decouples from its planet trap, entering into a phase of Type-II migration. The `gap opening' mass requires the planet's Hill radius to be larger than the pressure scale height at the planet's location, otherwise the gap will be closed by gas pressure. A second requirement is that the torques from the planet on the disk exceeds the torques caused by viscous stress. These requirements are summarized by,
(\cite{HP12},\cite{LP93}),\begin{align}
\frac{M_{gap}}{M_*} = min\left[3h_p^3,\sqrt{40\alpha_{turb} h_p^5}\right],
\label{eq:gapm}
\end{align}
where $h_p \equiv H_p/r_p$. When the gap is opened the planet decouples from the evolution of the planet trap, evolving on the viscous time\begin{align}
t_{mig} = t_{vis} = \frac{r_p^2}{\nu}.
\end{align}
This migration timescale is increased due to the inertia of the planet when it is above a critical mass $M_{crit} = \pi\Sigma_p a^2$, which is the approximate mass of the disk interior to the planet. Above the critical mass the migration time scale is,\begin{align}
t_{mig} = \frac{r_p^2}{\nu} (1 + M_p/M_{crit}),
\end{align}
and the radial evolution for Type-II migration is, \begin{align}
\frac{d r_{p.II}}{dt} &= \frac{r_p}{t_{mig}}.
\label{eq:t2}
\end{align}

Understanding how mass accretes onto the planet after the gap opens is still an open question and one that requires the results of detailed 3D numerical simulations of accretion onto migrating planetary cores (see for example the numerical simulation of \cite{K99}). It may be that the accretion is at least regulated by the global mass accretion rate in the disk. 

In this work, we parameterize the maximum mass of the planet with a fiducial value of fifty times the gap opening mass (fiducial value in \cite{HP12,HP13} and \cite{APC16a}), above which we shut off the accretion onto the planet. This is a necessary parameterization due to the unstable nature of the Kelvin Helmholtz rate of accretion. Between the gap opening mass and the maximum mass we maintain the same prescription for gas accretion as in Equation \ref{eq:acc01}.

\begin{figure}
\begin{center}
% \begin{overpic}[width=0.5\textwidth]{aug_plnt_track_smth.eps}
\begin{overpic}[width=0.5\textwidth]{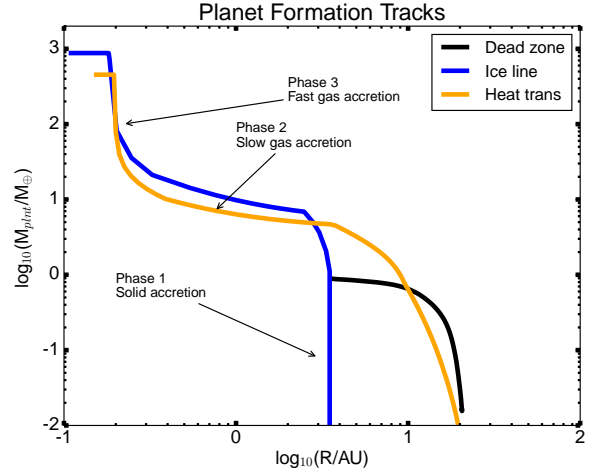}
\end{overpic}
\caption{Full planet tracks for our fiducial disk setup. Each track is associated with a single planet forming in isolation, without any dynamical interactions between other planets in other tracks. We also ignore interactions if two traps cross in space, so that there is no `trap jumping'. We annotate the three phases of mass accretion: Phase 1: Oligarchic growth, Phase 2: Slow gas accretion, limited by the Kelvin Helmholtz timescale and Phase 3: Rapid gas accretion, regulated by a Kelvin Helmholtz timescale $t_{KH} < O(10^5)$.}
\label{fig01}
\end{center}
\end{figure}

When combined with the planet trap model for planet migration this prescription for planet formation produces a well defined evolutionary track through the planet mass-semimajor axis space that we define as a `planet track'. 

In Figure \ref{fig01} we show three examples of such tracks. For our choice of disk parameters ($M_{disk}(t=0) = 0.1$ M$_\odot$, $s_{disk}(t=0) = 66$ AU, t$_{life} = 4.10$ Myr) we find that two of the planets end up being `Hot Jupiters' while the third fails to grow to a significant mass by the end of the disk lifetime. The location of these planet tracks, and the overall evolution of the growing planet depends on the location of the planet's natal planet trap. The location of the planet traps will depend on the disk model parameters that are selected, so a complete picture of possible planets will depend on a full population synthesis model that we leave for future work. Qualitatively, our planet tracks match the results from \cite{HP12}, apart from a trend that planets whose corotation torque saturates will tend to tilt farther to the left (migrate to lower orbital radii) during their phase of solid accretion than planets whose torques remain unsaturated.

During the oligarchic growth phase the mass accretion of the core is given by Equation \ref{eq:acc01b}. When the core reaches its isolation mass the mass accretion is dominated by gas accretion given by Equation \ref{eq:acc01c}. When the planet reaches its gap opening mass (Equation \ref{eq:gapm}) the radial evolution is no longer governed by Type-I migration, and is instead given by Equation \ref{eq:t2}.

Combining the core accretion model with planetary migration allows us to track the chemical composition of the material accreted onto the exoplanetary atmosphere. When combined with our full set of temporal snapshots of the chemical disk and our migration model outlined above, we can compute a full picture of the chemical makeup of the material being accreted onto the forming planet.

\section{ Atmospheric Compositions: Results }

We evolve a fiducial disk model with parameters listed in table \ref{tab:result01} and compute the formation of planets in each of the three planet traps outlined above. When evolved these disk model parameters produce a disk which ranges in mass from $0.013 \rightarrow 0.0044$ M$_\odot$ for a disk age range of $2\rightarrow 3$ Myr respectively. Submillimeter surveys of star forming regions have yielded mass ranges of $0.00143 - 0.0153$ M$_\odot$ for disks around stars with ages ranging between $2-3$ Myr (see \cite{And13}, \cite{Will13} and \cite{Cie15}). 

\subsection{ Condensation Front Locations: Comparison with TW Hya and HL Tau }

Some of the most important observational features in protoplanetary disks are condensation fronts (or equivalently `ice lines') which are the disk radii where the temperature falls below the sublimation temperature of volatile species. At these temperatures, the freezing efficiency of the volatile species approaches unity and the majority of chemical species become frozen onto grains. Carbon monoxide is a convenient molecule to study the effects of condensation fronts as it reacts with another easily detected molecule, N$_2$H$+$. In the gas phase, N$_2$H$+$ is created through protonation of gaseous N$_2$ down to a few degrees above the freezing temperature of CO ($\sim 20$ K). It is destroyed primarily through reactions with gaseous CO \citep{Qi13}. Another important observational tracer is formaldehyde (H$_2$CO) which can be efficiently produced by Hydrogen capture of frozen CO, then non-thermally desorbed to produce a gas component of H$_2$CO \citep{Qi13}. 

Observational surveys use the emission from these tracer molecules to infer the location of the CO condensation front. The quoted locations vary considerably between observational studies, from $25$, $30$ to $160$ AU (\cite{Pon14}, \cite{Qi13b} \& \cite{Qi13} respectively).
\begin{figure}
\centering
\includegraphics[width=0.5\textwidth]{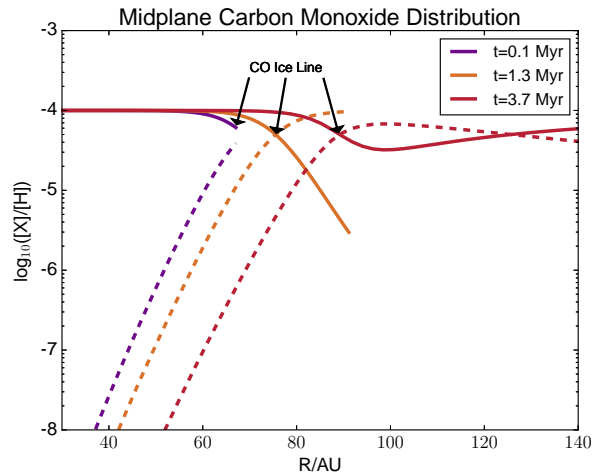}
\caption{Midplane CO distribution of the gas (solid line) component and frozen (dashed line) component, from 10 AU to 100 AU. The location of the CO ice line (condensation front) is denoted with the arrow, at a position that varies from $\sim$ 60 AU out to 65 AU for disk time from $0.1$ Myr to $3.7$ Myr respectively. }
\label{fig:numtest04}
\end{figure}

In Figure \ref{fig:numtest04} we show the ratio of CO gas and ice to Hydrogen atoms for radii ranging from $30$ AU to the outer edge of the disk, highlighting the evolution of the location of the CO ice line in our model. It evolves outward from $\sim$ 62 AU out to $\sim$ 90 AU between 0.1 and 3.7 Myr in the region of the disk that is primarily heated through direct irradiation. As seen in section \ref{sec:HTrap} the region of the disk that is heated through direct irradiation produces a temperature profile that is constant in time. Unlike water in Figure \ref{fig03}, carbon monoxide also evolves through a variety of other reactions other than freezing onto grains. In particular it is sensitive to the local UV field which photodesorbs the molecule off of dust grains. As the density of the dust is reduced the efficiency of CO absorption is reduced and the rate of photodesorption is increased. Both of these affects conspire to move the location of the ice line outward as the disk evolves and the density drops.

The condensation front of water is another important feature in protoplanetary disks as it represents a sharp transition in the opacity structure of the disk. Detecting the effect of the water ice line is generally limited by resolution, and only a limited number of detections have been completed with {\it Spitzer} and {\it Herschel}. In TW Hya a strong water vapour emission is detected at about 4 AU but drops off by orders of magnitude at 4.2 AU \citep{Z13}, suggesting that the ice line can be defined as a sharp transition point between the gas and ice phase of water. 

This sharp transition can be seen in our own work in Figure \ref{fig03} where the water vapour drops off by over two orders of magnitudes within half an AU. The disk around TW Hya is classified as a transitional disk, which is a class of old systems that are undergoing the final stages of photoevaporation of the gas disk. The potential detection of the ice line in TW Hya represents an upper limit to the radial  location of the ice line during the early stages of the protoplanetary disk. As the gas disk evaporates from the inside out, any remaining water ice will be photodesorbed back into the gas and then lost from the disk by the photoevaporating UV field of the host star. The strong water vapour detection would then represent the evaporation front of the disk as seen in theoretical calculations of photoevaporation \citep{Gort15}.

Condensation fronts have also been proposed as the source of the emission gaps observed in the HL Tau disk by ALMA. Recent experiments have suggested that icy covered grains tend to be `stickier', increasing the efficiency of dust coagulation. Because the (sub)mm observations of ALMA predominately detect the emission of micron sized grains, it fails to detect the larger grains that are expected to exist at the condensation front \citep{Zh15}. The observations made of HL Tau show three main bands of reduced intensity at approximately 13 AU, 32 AU and centered at 63 AU (although the reduced intensity is about 10 AU wide). 

\cite{Zh15} interpret the locations of the first two gaps as being the ice lines of water and pure or hydrated NH$_3$. While the third gap is interpreted as starting at the ice line of CO$_2$ and passing through the freeze out of CO and N$_2$. 

HL Tau is identified as a young disk, with an age $ O(10^5)$ yrs into the Class 2 phase, so we compare to our fiducial model at $t=10^5$ yr. The ice lines for H$_2$O, CO$_2$, NH$_3$, and CO are located in our fiducial model at approximately $3.5$, $10$, $20$, and $62$ AU respectively. We find that the approximate locations of the NH$_3$ and CO ice lines agree with the locations inferred through observations of HL Tau, which is the first source that could be studied in enough detail to infer these properties. The disagreement at small radii may imply that HL Tau is more representative of a disk with parameters that differ from our fiducial model. This region of the disk is heated primarily by viscous sources which tend to be more sensitive to disk parameters such as the initial mass accretion rate and disk size than the radiatively heated region at larger radii. Our forthcoming population synthesis paper will address these issues.

\subsection{ Computing the Initial Atmospheric Composition of a Forming Exoplanet }

We begin our planet formation model (see Table \ref{tab:result01}) at $0.1$ Myr, giving enough time for the initial core of $0.01$ M$_\oplus$ to grow and initialize our computation. At this initial time the three planet traps are located at:\begin{align}
{\rm Dead~zone:} &\quad 21 ~{\rm AU} \nonumber\\
{\rm Heat~transition:} &\quad 19.6~{\rm AU} \nonumber\\
{\rm Water~ice~line:} &\quad 3.5~{\rm AU} \nonumber
\end{align}
With the ice line and heat transition trap both starting within the dead zone we must assess whether the forming planet will actually be trapped using Equations \ref{eq:tlib} and \ref{eq:tvis}. Where the trap initially starts determines the length of time prior to the saturation of the corotation torque. The ice line saturates at a time of about $0.19$ Myr and the heat transition saturates later at $0.73$ Myr. 
\begin{figure}
\centering
\includegraphics[width=0.5\textwidth]{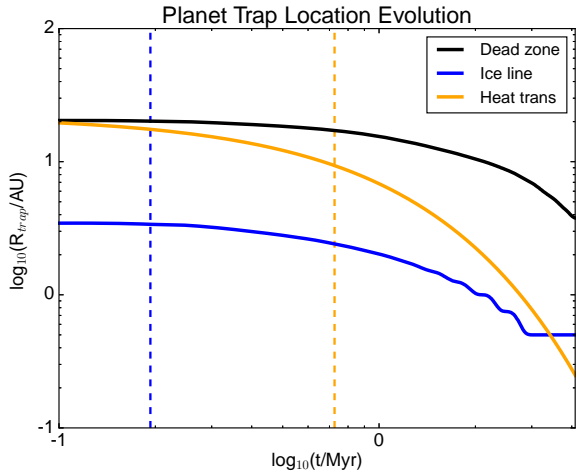}
\caption{The temporal evolution of the planet traps in our model. The vertical lines denote the location and time where the corotation torque saturates and the radial evolution of the planet decouples from the radial evolution of the trap.}
\label{fig:results01}
\end{figure}

\begin{figure}
\centering
\includegraphics[width=0.5\textwidth]{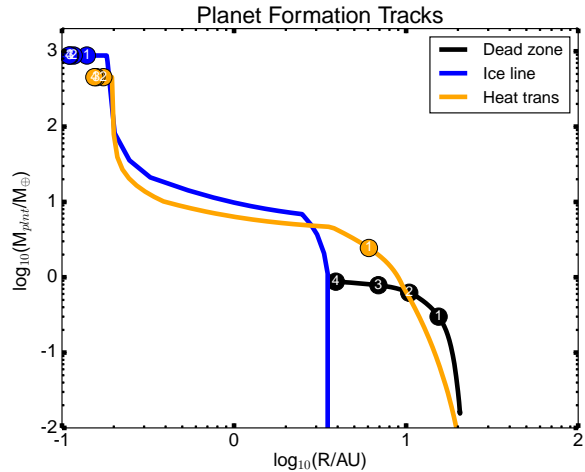}
\caption{Full planet formation tracks along with annotations showing the location of the planet in the mass-semimajor axis space for $1$, $2$, $3$ and $4$ Myr. At small radii planets spend a very short time accreting its solids and can reach super-Jovian masses in less than $1$ Myr. The saturation of the corotation torques also help to speed up formation of Jupiter mass planets, as the planet quickly moves into regions of the disk with higher surface densities.}
\label{fig:results02}
\end{figure}

Figure \ref{fig:results01} shows the time evolution of each planet trap and the location where the corotation torque saturates and the planet migrates under standard Type-I migration caused by Lindblad torques. When combined with our planet formation model we compute the planet tracks shown in Figure \ref{fig:results02}.

In Figure \ref{fig:results02}, each planet track is annotated to demonstrate how much time the planets linger in each evolutionary phase. In general we see that the planet which is initialized in the ice line trap grows very quickly, and reaches its maximum mass within $1$ Myr of disk evolution. This rapid growth is due to the rapid inward migration caused by the early saturation of the corotation torque at the location of the ice line. At larger radii, the heat transition planet accretes at a slower rate and takes $2$ Myr to reach its maximum mass. When the ice line and heat transition planets reach their maximum masses before the disk is dispersed, they proceed to evolve radially through Type-II migration for the remainder of the disk history. On the other hand, the dead zone planet does not reach its maximum mass by the time the disk disperses.

For the disk parameters shown in Table \ref{tab:result01} we compute three planets with the final planetary mass and orbital radius shown in Table \ref{tab:result02}. We compute two Hot Jupiters and one sub-Earth mass planet, the former having a small spread in planetary mass and orbital radius while the latter represents a `failed' core that never reached a sufficient mass to draw down an atmosphere. 
\begin{table}
\begin{center}
\caption{Final planetary properties at $t=t_{life}$}
\begin{tabular}{l c c}\hline
& $M ~(M_{{\rm Jupiter}})$ & $a~({\rm AU})$ \\\hline
Dead zone & $0.003 $ & $3.7$ \\\hline
Ice line & $2.67$ & $0.11$ \\\hline
Heat transition & $1.43$ & $0.15$ \\\hline
\end{tabular}\\
\label{tab:result02}
\end{center}
\end{table}

\begin{figure}
\centering
\includegraphics[width=0.5\textwidth]{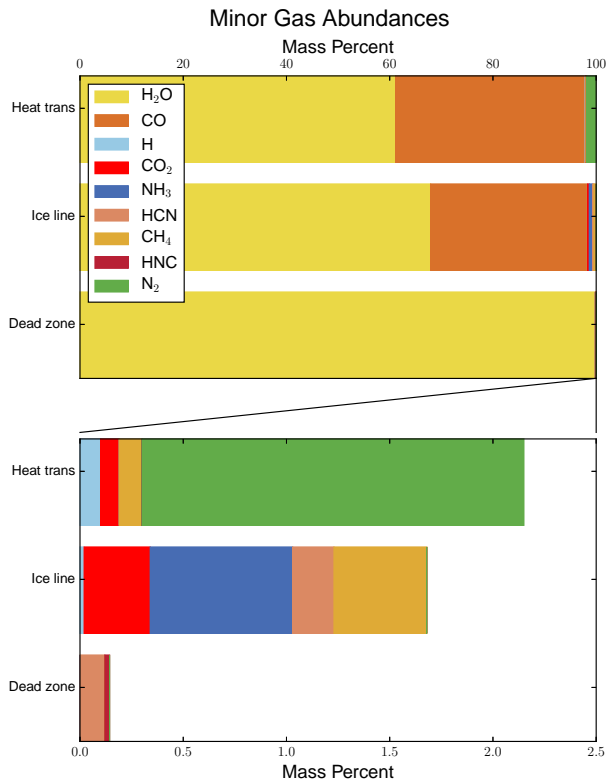}
\caption{ Mass percentage of most abundant gases other than H$_2$ and He. The bottom panel zooms into the gas species other than H$_2$O and CO. The heat transition and ice line planets show similar abundances in both H$_2$O and CO. The dead zone planet did not accrete an atmosphere, and the reported species are the most abundant ices that could contribute to an atmosphere through out gassing. It is exclusively dominated by water ice. The difference between the heat transition and ice line planets is in the nitrogen carriers, and is determined by {\it when} the planet undergoes unstable gas accretion (see Figure \ref{fig:results02}). At early times (when the ice line accretes) the nitrogen gas is primarily in NH$_3$ and HCN, while at later times  the disk cools and the nitrogen content is dominated by N$_2$. The values are shown in Table \ref{tab:app01}. }
\label{fig:results03}
\end{figure}

\begin{table}
\centering
\cprotect\caption{ Minor gas abundance as percent of mass not in H2 and He. The dead zone atmosphere is populated by a theoretical out gassing of frozen material. For a complete list of gas species please visit \verb+physwww.physics.mcmaster.ca/~cridlaaj/Format_atmos_data.html+ }
\begin{tabular}{ l c c c }\hline
& Dead zone & Ice line & Heat transition \\\hline
H2O & $99.84$ & $67.92$ & $61.15$ \\\hline
CO & $< 0.01$ & $30.40$ & $36.70$ \\\hline
H & $< 0.01$ & $0.02$ & $0.1$ \\\hline
CO2 & $< 0.01$ & $0.32$ & $0.09$ \\\hline
N2 & $< 0.01$ & $< 0.01$ & $1.85$ \\\hline
NH3 & $< 0.01$ & $0.69$ & $< 0.01$ \\\hline
HCN & $0.12$ & $0.2$ & $< 0.01$ \\\hline
CH4 & $< 0.01$ & $0.45$ & $0.11$ \\\hline
HNC & $0.02$ & $< 0.01$ & $< 0.01$ \\\hline
\end{tabular}
\label{tab:app01}
\end{table}

In Figure \ref{fig:results03} we show the cumulative minor gas abundances as a percentage of the total mass of each planet formed in the three planet traps. These minor gas abundances are all gas species other than H$_2$ and He (which dominate the total mass of the planet, making up between $99.5\%$ and $99.8\%$ of the total mass). Only the gas species that have mass percentages greater than $0.01\%$ of the total minor gas mass are shown (see Table \ref{tab:app01} for numerical values). The dominant minor gases in our planets are CO and H$_2$O, owing to the high abundance of these molecules in the disk. For the dead zone planet which accreted no atmosphere, we have estimate an `out gassed' atmosphere based on the amount of frozen material it accreted during the oligarchic growth phase. In producing Figure \ref{fig:results03} we have assumed that all of the ice that is accreted is out gassed into an atmosphere.

The ratio between H$_2$O and CO in the atmospheres of gas giants is determined largely by the relative amount of those species in the gas phase where the planet rapidly draws down the majority of its gas (Phase 3 accretion). If this occurs outside the radial location of the ice line the majority of the water is in the ice phase rather than the gas phase. The dead zone planet underwent its oligarchic growth outside of the water ice line and within the CO ice line, hence its high abundance of water in its hypothetical atmosphere.

By contrast the abundance of the nitrogen carriers (primarily NH$_3$, N$_2$, HCN) show variation between the ice line planet (primarily NH$_3$ and HCN) and the heat transition planet (primarily N$_2$). This variation appears to depend on {\it when} the gas giant accretes its gas. As seen in Figure \ref{fig:results02} the ice line accretes its gas very early in the disk evolution (prior to $1$ Myr) while both the heat transition and dead zone accrete later than $1$ Myr. As the disk ages the viscously heated regime of the disk cools and the gas phase nitrogen content of the disk is primarily converted to N$_2$ while species like NH$_3$ and HCN freeze out onto grains. 

Figure \ref{fig:results03} assumes that no chemical processing has occurred once the gas has accreted onto the planet. We call these atmospheres {\it early} atmospheres as they represent the chemical abundances of a homogeneous, passive atmosphere that has not had time to change the chemical abundances that it accretes during its formation. In what follows, when we compare the chemical results of our theoretical atmospheres to the observational data we also assume that there is no physical processing (vertical transport, atmospheric loss) of the atmosphere to change its chemical make up.

\begin{figure*}
\centering
\includegraphics[width=\textwidth]{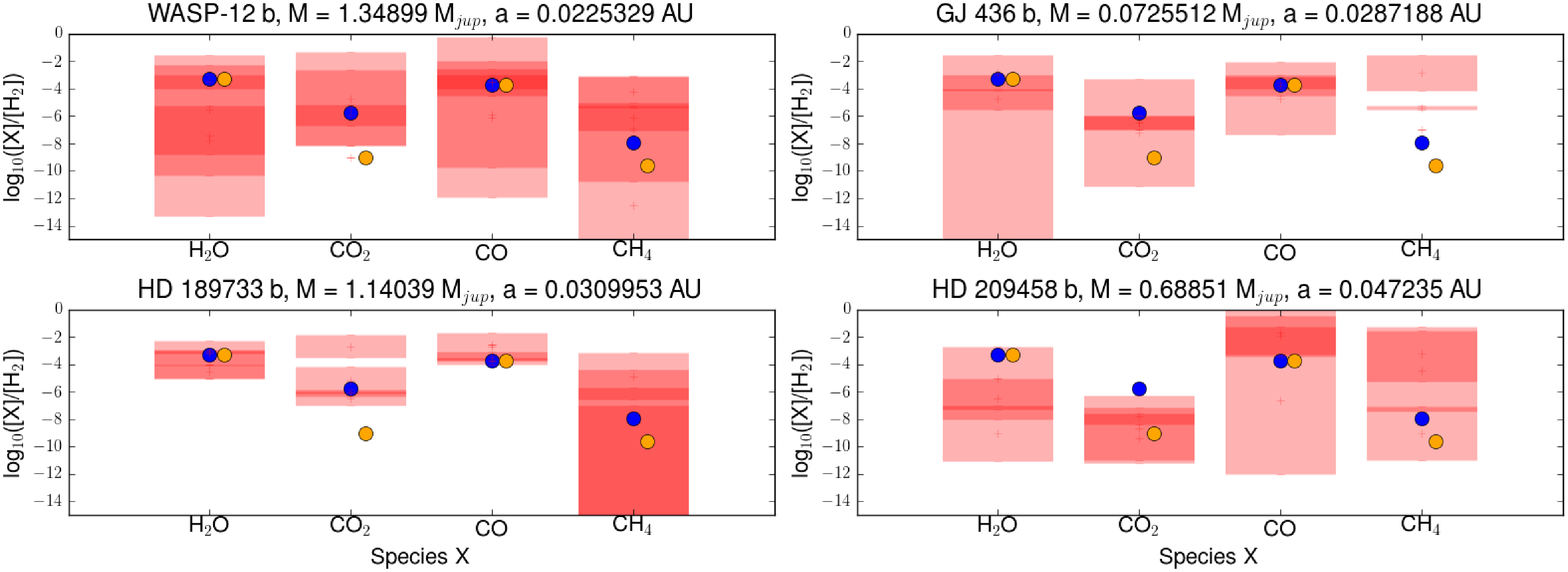}
\caption{ Mixing ratio $[X]/[H_2]$: the abundance of the molecule X over the abundance of molecular Hydrogen. The red bars denote the range of inferred mixing ratios from \citet{MadSea09,MadSea10,MigKal13,Lee11,Lin11,Lin13}, where the darker opacities denote regions of overlap between different sources. The mixing ratios are shown in the colour denoted in Figure \ref{fig:results02}: Ice line (blue) and Heat transition (orange). The dead zone planet is absent from this comparison as it is not representative of a Hot Jupiter atmosphere. }
\label{fig:results04}
\end{figure*}

\section{ Discussion }
\subsection{ Comparison with the Observations }

We wish to compare the computed chemical abundance of the atmospheres of our theoretical planets to the available retrieved abundances from observed planets. The atmospheric data has generally been obtained for Hot Jupiters with orbital seperations of a few hundredths of an AU (\cite{Lee11}, \cite{Lee13}, \cite{Lin11}, \cite{Lin13}, \cite{MadSea09}, \cite{MadSea10} and http://exoplanets.eu for orbital properties). The orbital radii of the ice line and heat transition planets (R$\sim 0.11, 0.15$ respectively) are an order of magnitude farther out than the planets available in the atmospheric data. Regardless, we will compare their atmospheric contents to the ones in the data because our theoretical planets could migrate into the orbital radii range of Hot Jupiters if the disk was allowed to evolve for longer than 4.1 Myr. Moreover, we assume that the chemical content of the atmosphere is unchanged once the planet reaches its maximum mass, because it no longer 
accretes any disk material and we do not evolve the atmosphere once it has been accreted onto the planet. Indeed, in a full population synthesis model one of the variable parameters is the lifetime of the disk, which can have a range of values as low as 1 Myr and as high as at least 10 Myr. A longer disk lifetime would also let the dead zone planet accrete more matter which would allow it to collect an atmosphere. For this reason, the disk lifetime is one of the most important disk parameters to study in future work.

Figure \ref{fig:results04} shows the range of inferred mixing ratios from observations (see \citet{MadSea09,MadSea10,MigKal13,Lee11,Lin11,Lin13}) and compares the range to our theoretically computed mixing ratios. 

Generally speaking we see a good agreement for our computed models in the ice line and heat transition trap as well as the range of inferred ratios that had the highest agreement (darkest region) between observers. While we lack the statistical breadth of a full parameter search to fully understand the connection between our theoretical predictions and the observations, these results suggest that our early (non-processed) atmospheres are a good representation of the chemical make up of observed Hot Jupiter atmospheres. Variations between these early atmospheres and the observations have a variety of potential causes. This may include chemical processing in the atmosphere as well as mixing due to turbulent transport and atmospheric loss from a variety of sources. Only the chemical abundance of the upper atmosphere (down to one optical depth) of the planet can be observed through emission and transmission spectra. Thus the observable abundance of CH$_4$ and H$_2$O is sensitive to UV photolysis, while H$_2$O 
and CO$_2$ are sensitive to the efficiency of vertical mixing. Alternatively, molecules like CO appear to be unchanged when testing different UV flux and mixing efficiency \citep{MigKal13}.

\subsection{ Elemental Ratios: C/O }

The elemental ratio of an atmosphere is another probe for studying the chemical abundance and physical properties of exoplanetary atmospheres. The most relevant elements are Carbon and Oxygen because of their abundance and ease of detection. Nitrogen is a third important element however it has yet to be directly detected or inferred in the emission spectra of Hot Jupiters. The carbon-to-oxygen ratio ($C/O$) is a widely studied elemental ratio and is believed to track the approximate formation location of gas giant planets \citep{O11}. 

The general picture assumes that the abundance of H$_2$O, CO$_2$ and CO are only determined by their condensation onto grains at each of their respective ice lines. If a gas giant accretes its gas inside the CO ice line but outside the ice lines of both CO$_2$ and H$_2$O then $C/O\sim 1$. If the planet accretes within the CO$_2$ ice line then $0.66 < C/O < 1$ depending on the relative abundance of CO$_2$ to CO with the lowest ratio requiring equal amounts of CO$_2$ and CO. CO$_2$ is rarely more abundant than CO, as is the case for the chemistry outputs shown here. Within the water ice line a planet would accrete an elemental ratio $C/O \sim 0.5$ if H$_2$O and CO are equally abundant and CO$_2$ is negligible (as we see in our planets) and $C/O\sim 0.25$ if H$_2$O, CO$_2$ and CO are all equally abundant. 

In our model the ice line and heat transition planets have $C/O = 0.227; 0.279$ respectively. The dead zone planet has an incredibly low $C/O \sim 0.001$ in its out gassed atmosphere. Both the heat transition and ice line planets have C/O of less than a third, due to the abundance of water over CO and the low abundance of CO$_2$. The difference in relative abundances and C/O appears to arise because the heat transition trap begins its gas accretion outside of the ice line, before moving within the location of the water condensation front where it accretes the vast majority of its gas.  The planet that started forming in the water ice line began and ended its gas accretion within the ice line due to the early saturation of the corotation torque in the water ice line. Because of its early saturation, the ice line accretes more water (by mass) than the heat transition trap.

Again we see the importance of {\it where} and {\it when} the planet forms: the radial location of gas accretion sets the bulk of the carbon-to-oxygen ratio. We initialized our chemical code with the dominant chemical species computed from a chemical simulation of a molecular cloud. This initial condition has a C/O of 0.288 which is close to the C/O ratio of both planets. This suggests that the gas that each planet accreted had very little elemental processing prior to accretion. This processing depends on the physical conditions of the disk, and thus is sensitive to the radial location of the forming planet and the age of the disk.

\subsection{ Elemental Ratios: C/N }

The Carbon-to-Nitrogen ratio has yet to be used for analyzing exoplanetary atmospheres. However it has been used to characterize the astronomical bodies in our solar system (for example, the review by \cite{Berg15}). Our sun has C/N $\sim 4$ while the Earth has an average ratio of $\sim 49$. This large discrepancy is likely caused by the Earth accreting material that had been thermally processed \citep{Berg15}. For the Hot Jupiter formed in the ice line trap we find C/N $\sim 47$. For the Hot Jupiter formed in the heat transition trap we find C/N $\sim 10$. Drawing the same conclusion as before, this discrepancy is caused by the thermal processing of the gas disk. This thermal processing is reduced as the disk cools, so C/N can be used as a measure of {\it when} the planet accretes its gas. High ratios suggest that the planet forms earlier in the disk lifetime. The ice line planet began its gas accretion at about 0.29 Myr into the simulation while the heat transition planet began accreting gas at 1.13 Myr. 
From Figures \ref{fig:gaschem} and \ref{fig:icechem} we can see that a significant amount of chemical evolution has taken place between these two starting points.

\section{ Conclusions }

We have developed a method of linking the time dependent, non-equilibrium astrochemistry of accretion disks to the chemical abundances of exoplanetary atmospheres. With this model of a disk undergoing viscous evolution as well as photoevaporation, we have run $4.1$ Myr lifetime model of a fiducial disk and planet formation model to produce a set of exoplanets in three basic planet traps. We find that:
  \begin{itemize}
  \item It contains condensation fronts of H$_2$O, CO$_2$, NH$_3$ and CO located at $3.5$, $10$, $30$ and $62$ AU at $t= 0.1$ Myr respectively, corresponding well with the inferred locations of the condensation fronts proposed to exist in HL Tau .
  \item We compute three planets with M = 2.67 M$_{jup}$, 1.43 M$_{jup}$ and 0.003 M$_{jup}$ and a = 0.11 AU, 0.15 AU and 3.7 AU in the ice line, heat transition and dead zone traps respectively.
  \item In these three traps we find an atmospheres with Carbon-to-Oxygen ratios (C/O) of 0.227, 0.279, 0.001 in the ice line, heat transition and dead zone traps respectively .
  \item Three planets were produced that had mixing ratios for CO, H$_2$O, CO$_2$ and CH$_4$, of $1.99\times 10^{-4}$, $5.0\times 10^{-4}$, $1.8\times 10^{-6}\rightarrow 9.8\times 10^{-10}$ and $1.1\times 10^{-8}\rightarrow 2.3\times 10^{-10}$ respectively. These correspond well with the mixing ratios inferred from observed emission spectra.
  \item C/O is largely influenced by the location of the accreting protoplanet with respect to the location of the water ice line when the planet accretes the majority of its gas.
  \item C/N is influenced by thermal processing of the accreted gas and is set by the location of the accreted gas and when the gas was accreted.
  \item The chemical abundance of the atmosphere depends on {\it where} and {\it when} the planet accretes its atmosphere. This is seen in both the carbon-to-oxygen ratio as well in the mass weighted abundances.
  \item Early (in time) accreters generally have their nitrogen content in the form of NH$_3$ and HCN, while the later accreters have their nitrogen in the form of N$_2$.
  \end{itemize}

It is not clear if these results are representative of the full population of exoplanets, or are simply the result of our very limited sample of disk parameters. In future work, we will improve our methods by including dust growth, fragmentation, settling and radial transport into the radiative transfer model which will change the global ionization structure of the disk. Additionally we will perform a population synthesis study based on distributions of disk lifetimes and initial masses in order to develop a statistical treatment of planetary atmospheres.

\section*{ Acknowledgments }

We thank an anonymous referee for useful comments that helped to improve the paper. We are grateful to Ted Bergin for the use of his non-equilibrium code. We benefited from many discussions with him and his (former) graduate student Ilse Cleeves on astrochemical codes and analysis. Additionally we thank Til Birnstiel, Roy van Boekel, Cornelis Dullemond, Thomas Henning, Hubert Klahr, Oliver Gressel and Dmitry Semenov for interesting discussions and feedback on this work. The work made use of the Shared Hierarchical Academic Research Computing Network (SHARCNET: www.sharcnet.ca) and Compute/Calcul Canada. A.J.C. acknowledges funding from the National Sciences and engineering Research Council (NSERC) through the Alexander Graham Bell CGS/PGS Doctoral Scholarship. R.E.P. is supported by an NSERC Discovery Grant. M.A. acknowledges funding by an Ontario Graduate Scholarship (OGS) and NSERC CGS-M scholarship. R.E.P. also thanks the MPIA and the Institut f¨ur Theoretische Astrophysik (ITA) in the Zentrum f¨ur 
Astronomie Heidelberg for support during his sabbatical leave (2015/16) during the final stages of this project. A.J.C also thanks MPIA and the Institut f¨ur Theoretische Astrophysik (ITA) in the Zentrum f¨ur Astronomie Heidelberg for their hospitality during his 1 month stay in 2016.

%\newpage
\section*{Appendix}
\appendix
%\onecolumn
\section{ Viscous Evolution of Mass }\label{sec:calcmod}

Solving the eqs. \ref{eqn14} and \ref{eqn19} we find that the viscously evolving mass evolves as a power law. When there is no radiative region present, the mass evolution takes the form: \begin{align}
M(t) = \frac{M_0}{\left(1 + t/\tau_{vis}\right)^{3/16}},
\label{eq:app01}
\end{align}
where \begin{align}
\tau_{vis} = \frac{3 M_0}{16 \mdot_0}.
\label{eq:app02}
\end{align}
When the radiative regime appears (at a time of t = t$_1$) the mass evolves as:\begin{align}
M(t) = \frac{M_1}{\left[1 + (t-t_1)/\tau_{rad}\right]^{7/13}},
\label{eq:app03}
\end{align}
where \begin{align}
\tau_{rad} = \frac{7 M_1}{13 \mdot_1},
\label{eq:app04}
\end{align}
where M$_1 \equiv $ M$(t=t_1)$ from eq. \ref{eq:app01}.

These equations describe how the total mass of the disk would evolve if photoevaporation was neglected. They are also used in deriving the radial expansion of the disk through angular momentum transport (eq. \ref{eqn13}).

\begin{figure}
\centering
\includegraphics[width=0.5\textwidth]{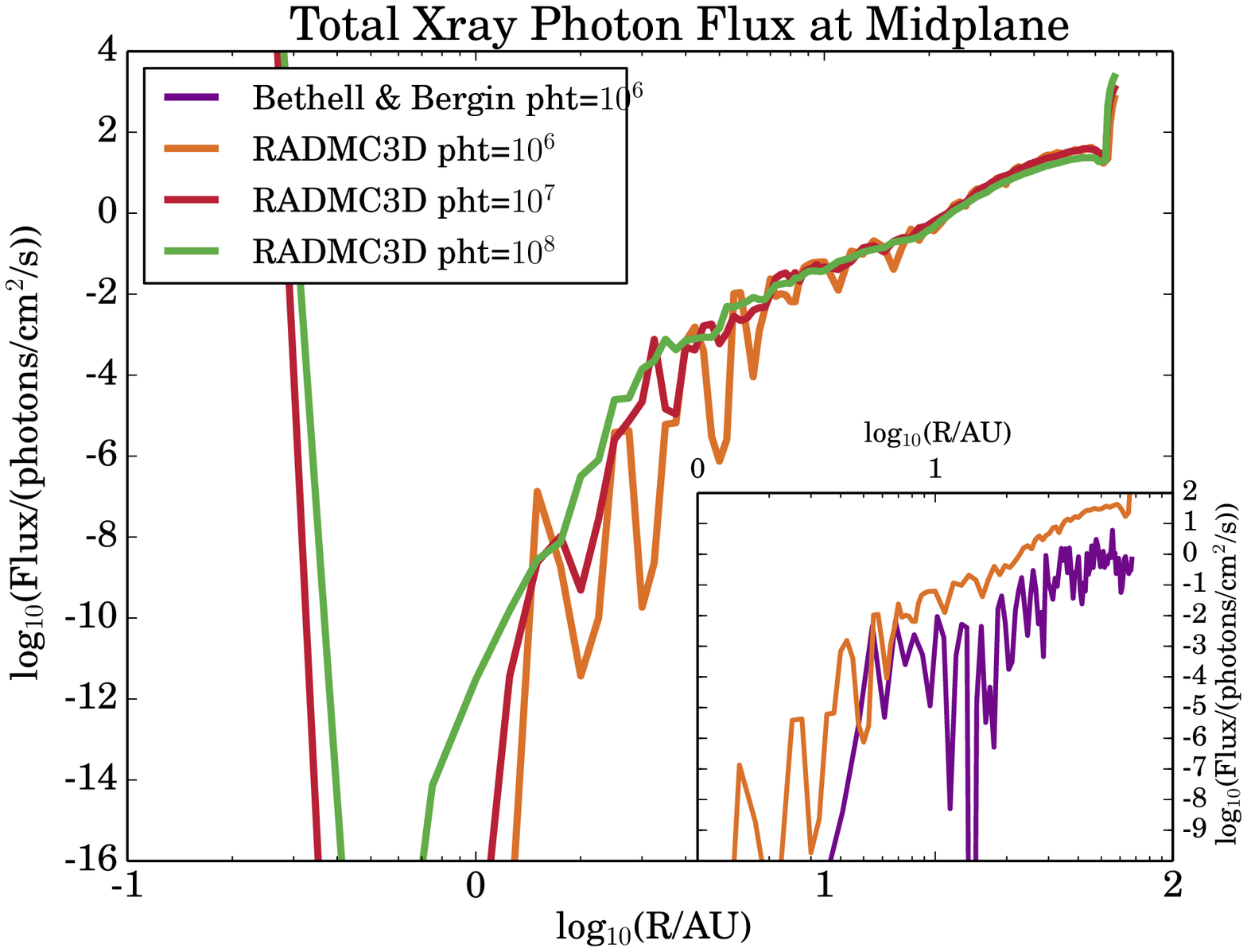}
\caption{Comparison of the flux of photons at the midplane in the fiducial disk. The three examples from RADMC3D show a rapid convergence in the flux at high radii, only showing a small deviation between the three photon resolutions. We compare the Bethell and Bergin method, which included a more accurate representation of the local molecular Hydrogen number density, to RADMC3D at the same level of photon number. These curves show similar levels of variation at low radii and an overall decrease in flux at far radii, caused by the more accurate Hydrogen number density profile.}
\label{fig:numtest01}
%\end{figure}
%\begin{figure}
%\centering
\includegraphics[width=0.5\textwidth]{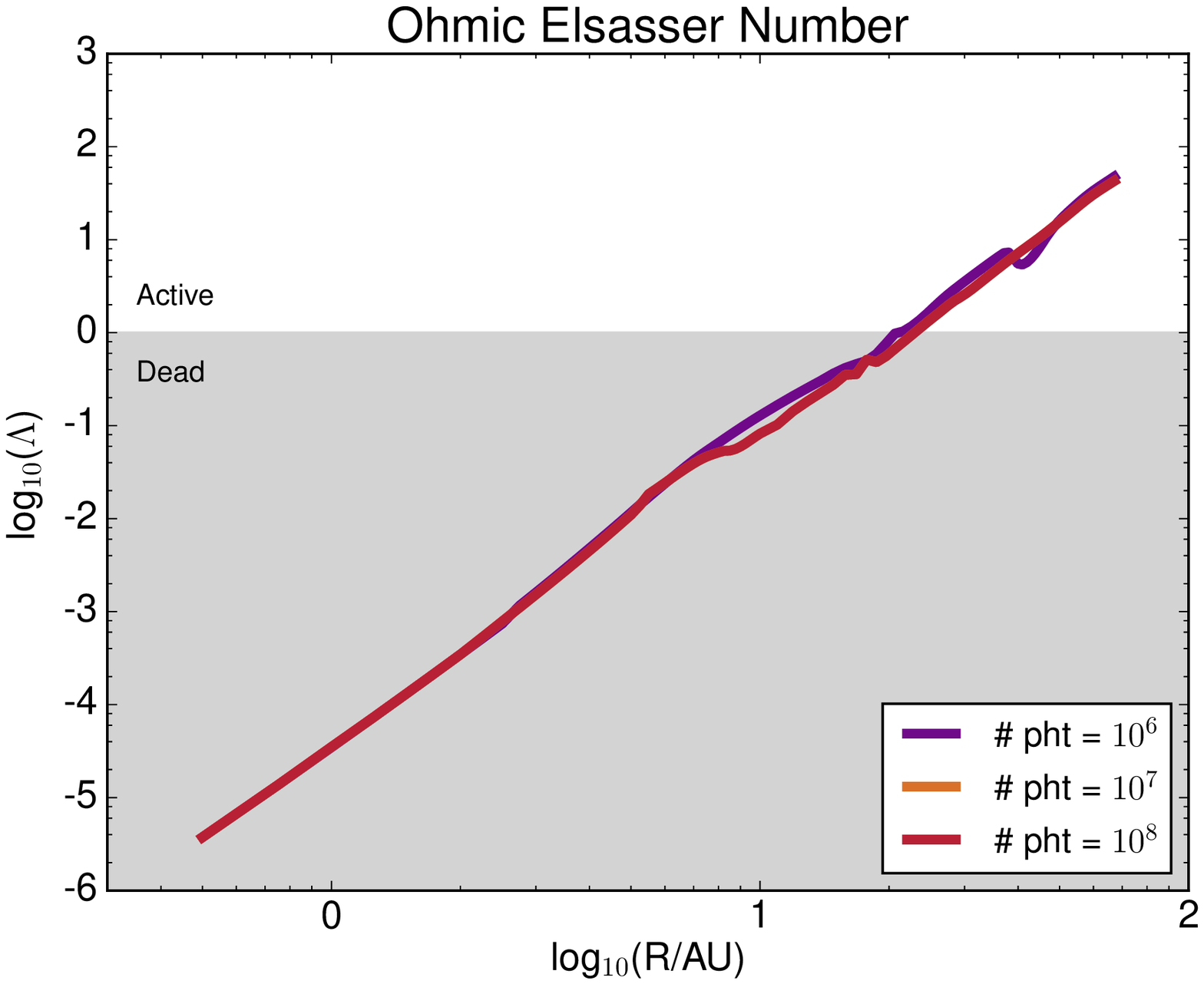}
\caption{Comparison of the resulting Ohmic Elsasser number (equation \ref{eq:in02}) from radiative transfer calculations with varying photon packet number. While Figure \ref{fig:numtest01} shows some variation in the photon count at small radii on the midplane, it does not translate into a large change in the ionization structure and the resulting Ohmic Elsasser number.}
\label{fig:numtest02}
\end{figure}

\section{ Numerical Tests and Verification }

Here we show tests of our numerical methods that have been used in this work. Our choice of using RADMC3D code in order to do the radiative transfer calculations for disk astrochemistry was motivated by our requirement for a computationally efficient method of determining the radiation field throughout the entire evolution of the disk over a few million years. 

In Figure \ref{fig:numtest01} we show a comparison of the total X-ray photon flux between RADMC3D at three different photon resolutions and the radiative transfer scheme based on the work by \cite{BB11x,BB11}. Generally RADMC3D quickly converges, showing only small variations at high radii but larger variations at smaller radii for different photon resolutions. The Bethell and Bergin code shows a similar radial dependence of the midplane photon flux, however generally has lower flux than RADMC3D over all radii. The \cite{BB11x,BB11} code computes the number density of molecular Hydrogen and the resulting Lyman-alpha flux iteratively with a Monte Carlo radiative transfer scheme. This iterative process results in a radiation field and molecular gas distribution that is a self-consistent, but increases computation time and results in higher variation in midplane flux between successive radii. Conversely we do not update the distribution of molecular hydrogen caused by the radiation field that is computed by 
RADMC3D. This results in a much faster computation, but will ultimately underestimate the flux of Lyman alpha photons along the midplane. Because of our desired temporal resolution over the lifetime of the disk we accept this trade off.

In Figure \ref{fig:numtest02} we show that using different photon resolution does not result in large differences in the ionization structure and therefore the radial dependence of the Ohmic Elsasser number. 

\begin{figure*}
\centering
\includegraphics[width=\textwidth]{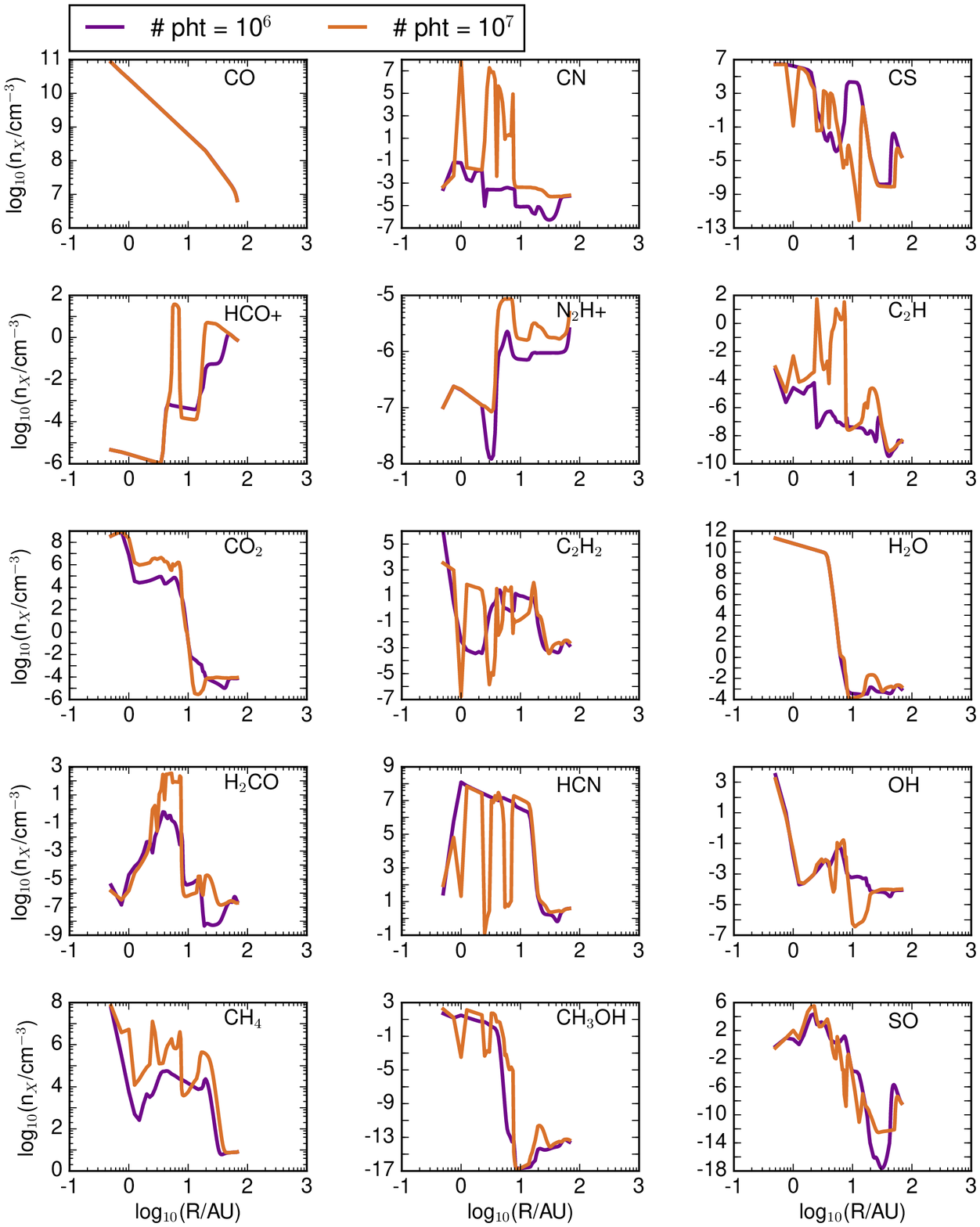}
\caption{Comparison of resulting gas abundances from different photon packet counts for the fiducial disk at $t = 10^5$ yrs. Small variation is seen in the most abundant species, while some of the more rare species that depend on the local radiation field are most sensitive to resolution. The results from the $10^7$ and $10^8$ packet tests are identical, so the result of $10^8$ packets is not shown.}
\label{fig:numtest03}
\end{figure*}

In Figure \ref{fig:numtest03} we show that the chemical abundances of important gas species resulting from the three different photon resolutions have very little radial differences. There is virtually no difference in the results obtained from using resolution of $10^7$ and $10^8$ photons with RADMC3D. There is however some difference in the resulting abundances of species that are closely dependent on the local radiation field. For example the gas species of CN is produced through the UV photolysis of HCN, and so the ratio of the abundance of these species are generally used as an observational tracer for the strength of the UV field in accretion disks. In comparing the two photon resolutions, it is clear that the lower resolution calculation is filtering out some large variations that happen over very small regions of the disk. However the global properties are generally maintained for the most abundant molecules in the disk. Additionally, for molecules that have been reportedly observed in planetary 
atmospheres, H$_2$O, CO, CO$_2$ and CH$_4$, only CH$_4$, which has the highest observational uncertainty shows significant variations in abundances between the two photon resolutions. For this first run of the model we will use the lower resolution, $10^6$ photon packets, in computing the chemistry.

\subsection{ Observational Tracers }

Observational tracers of the chemical structure of protoplanetary disks are obtained primarily through the observations of gas emission spectra in the millimeter to submillimeter range. Due to observational constraints, the average excitation temperature and total surface density of a molecule are the primary properties that can be determined through spectral observations. To compare to available observations, we have computed the total surface density of a given molecular species. We have assumed that the disk is viewed face on and that the observed line intensities are due to a cumulative effect of all the molecules of a given species, spread out over the total surface area of the disk. The total column number density therefore has the form:\begin{align}
N_{tot}&= \frac{1}{\pi s^2}\int n_X(R,z) R dRd\theta dz \quad {\rm cm}^{-2},
\label{eq:numtest01}
\end{align}
where $n_X$ is the volume number density in units of cm$^{-3}$ and $s$ is the outer edge of the disk at a given time. The total surface density relative to the total surface density of carbon monoxide is shown in Table \ref{tab:numtest01}. 

In general we find a fair agreement between the ratios of surface density for many of the detected molecules, with variations on the order of unity in H$_2$O, HCO+, CH$_4$ and H$_2$CO. For other molecules like C$_2$H, CN and C$_2$H$_2$ we underestimate the ratio by a few orders of magnitude, likely caused by radiative transfer effects like optical depth. For an optically thick line emission, its intensity does not represent the full line of sight surface area of the gas. For the case of $^{12}$CO, its line emission is generally optically thick. Therefore, in using $^{12}$CO as the denominator of the ratios that make up Table \ref{tab:numtest01}, observations are at risk of over estimating the true ratio between a molecule and $^{12}$CO. This depends on whether the line emission of the molecule that makes up the numerator is optically thin, or if the optical depth of the line is drastically different than the optical depth of $^{12}$CO.

\begin{table}
\begin{center}
\caption{ Total surface number density relative to $^{12}$CO at three instances in the disk as computed with our chemistry model. Corresponding ratios from the results of multiple observational sources are included in the right-most column for comparison.}
\label{tab:numtest01}
\begin{tabular}{l c c c c}\hline
& $0.1$ Myr & $1.3$ Myr & $3.7$ Myr & Literature\\\hline
H$_2$O & $1.09(-2)$ & $6.59(-4)$ & $1.01(-4)$ & $5.09(-2)$ $^c$\\\hline
OH & $4.90(-6)$ & $6.72(-6)$ & $1.85(-5)$ & $1.46(-2)$ $^c$\\\hline
CO$_2$ & $8.40(-6)$ & $2.16(-6)$ & $7.03(-7)$ & \\\hline
CH$_4$ & $4.76(-5)$ & $8.22(-5)$ & $6.57(-5)$ & $2.47(-4)$ $^c$\\\hline
HCO$+$ & $3.58(-7)$ & $8.29(-7)$ & $1.64(-6)$ & $4.78(-5)$ $^b$\\\hline
C$_2$H & $1.44(-5)$ & $2.53(-5)$ & $4.29(-5)$ & $9.50(-2)$ $^a$\\\hline
CN & $1.55(-5)$ & $3.22(-5)$ & $7.33(-5)$ & $1.71(-3)$ $^a$\\\hline
HCN & $5.60(-4)$ &  $7.37(-4)$ & $2.12(-3)$ & $3.59(-5)$ $^a$\\\hline
H$_2$CO & $5.99(-6)$ & $1.19(-5)$ & $8.58(-6)$ & $8.81(-6)$ $^b$\\\hline
C$_2$H$_2$ & $1.96(-6)$ & $3.94(-6)$ & $1.64(-6)$ & $2.21(-3)$ $^c$\\\hline
N$_2$H$+$ & $2.86(-9)$ & $5.51(-9)$ & $7.23(-7)$ & \\\hline
CS & $2.28(-6)$ & $3.68(-6)$ & $1.74(-5)$ & \\\hline
CH$_3$OH & $1.78(-11)$ & $1.62(-12)$ & $6.24(-13)$ & \\\hline
\end{tabular}
\end{center}
{\flushleft
Note: format $a(b)$ denotes $a\times 10^b$\\
$^a$ Averaged values from table 6 of \cite{Kast14} assuming the ratio $^{12}$CO$/^{13}$CO$ = 77$ \citep{Wal13}\\
$^b$ Averaged values from table 4 of \cite{Pun15} assuming the ratio $^{12}$CO$/^{13}$CO$ = 77$ \\
$^c$ Values taken from table 3 of \cite{Mand12} and represents line ratios based on a slab model \\
}
\end{table}

\ignore{
\section{ Minor Gas Abundance }

Numerical values from figure \ref{fig:results03}. In selecting the plotted chemicals only species with at least one percent (by mass) was shown. All upper limits show species that were detected in some planets but not others.

\begin{table}
\centering
\caption{ Minor gas abundance as percent of mass not in H2 and He }
\begin{tabular}{ l c c c }\hline
& Dead zone & Ice line & Heat transition \\\hline
H2O & $99.84$ & $67.92$ & $61.15$ \\\hline
CO & $< 0.01$ & $30.40$ & $36.70$ \\\hline
H & $< 0.01$ & $0.02$ & $0.1$ \\\hline
CO2 & $< 0.01$ & $0.32$ & $0.09$ \\\hline
N2 & $< 0.01$ & $< 0.01$ & $1.85$ \\\hline
NH3 & $< 0.01$ & $0.69$ & $< 0.01$ \\\hline
HCN & $0.12$ & $0.2$ & $< 0.01$ \\\hline
CH4 & $< 0.01$ & $0.45$ & $0.11$ \\\hline
HNC & $0.02$ & $< 0.01$ & $< 0.01$ \\\hline
\end{tabular}
\label{tab:app01}
\end{table}
}

\label{lastpage}


\begin{thebibliography}{99}
\bibitem[\protect\citeauthoryear{Aikawa \& Herbst}{1999}]{AH99} Aikawa Y. \& Herbst E. 1999. A\& A {\bf 351}: 233.

\bibitem[\protect\citeauthoryear{Alessi, Pudritz \& Cridland}{2016a}]{APC16a} Alessi M., Pudritz R.E. \& Cridland A.J. 2016a. submitted.

\bibitem[\protect\citeauthoryear{Alessi, Pudritz \& Cridland}{2016b}]{APC16b} Alessi M., Pudritz R.E. \& Cridland A.J. 2016b. in prep.

\bibitem[\protect\citeauthoryear{Alexander et al.}{2014}]{Alex14} Alexander R., Pascucci I., Andrews S., Armitage P. \& Cieza L. 2014. Protostars and Planets VI. The University of Arizona Press.

\bibitem[\protect\citeauthoryear{Alibert et al.}{2005}]{Alb05} Alibert Y., Mordasini C., Benz W. \& Winisdoerffer C., 2005, A\& A {\bf 434}: 343.

\bibitem[\protect\citeauthoryear{Alonso et al.}{2004}]{Alo04} Alonso R., Brown T.M., Torres G., Latham D.W., Sozzetti A., Mandushev G., Belmonte J.A., Charbonneau D., Deeg H.J., Dunham E.W., O'Donovan F.T \& Steganik R.P. 2004. ApJ {\bf 613}: L153-L156.

\bibitem[\protect\citeauthoryear{Andrews et al.}{2013}]{And13} Andrews S.M., Rosenfeld K.A., Kraus A.L. \& Wilner D.J. 2013. ApJ {\bf 771}: 129.

\bibitem[\protect\citeauthoryear{Bai \& Stone}{2013}]{BS13} Bai X-N. \& Stone J.M. 2013 {\bf 769}: 76.

\bibitem[\protect\citeauthoryear{Baillie \& Charnoz}{2014}]{BC14} Baillie K \& Charnoz S. 2014 ApJ {\bf 786}: 35.

\bibitem[\protect\citeauthoryear{Baillie, Charnoz \& Pantin}{2015}]{BCP15} Baillie K, Charnoz S. \& Pantin E. 2015 A\& A {\bf 577}: A65.

\bibitem[\protect\citeauthoryear{Balbus \& Hawley}{1991}]{BH91} Balbus S.A. \& Hawley J.F., 1991, ApJ {\bf 376}: 214.

\bibitem[\protect\citeauthoryear{Batalha et al.}{2013}]{Kep12} Batalha N.M, Rowe J.F, Bryson S.T., Barclay T. et al. 2013 ApJS {\bf 204}: 24.

\bibitem[\protect\citeauthoryear{Bell \& Lin}{1994}]{BL94} Bell K.R. \& Lin D.N.C. 1994. ApJ {\bf 427}: 987.

\bibitem[\protect\citeauthoryear{Bell et al.}{1997}]{Bea97} Bell K.R., Cassen P.M., Klahr H.H. \& Henning T. 1997. ApJ {\bf 486}: 372.

\bibitem[\protect\citeauthoryear{Bergin et al.}{2003}]{Berg03} Bergin E., Calvet N., D'Alessio P. \& Herczeg G.J. 2003, ApJ {\bf 591}: L159.

\bibitem[\protect\citeauthoryear{Bergin et al.}{2015}]{Berg15} Bergin E., Blake G.A., Ciesla F., Hirschmann M.M. \& Li J. 2015. PNAS {\bf 112}: 8965.

\bibitem[\protect\citeauthoryear{Bethell \& Bergin}{2011a}]{BB11x} Bethell T. \& Bergin E. 2011a. ApJ {\bf 739}: 78.

\bibitem[\protect\citeauthoryear{Bethell \& Bergin}{2011b}]{BB11} Bethell T. \& Bergin E. 2011b. ApJ {\bf 740}: 7.

\bibitem[\protect\citeauthoryear{Birnstiel et al.}{2012}]{Birn12} Birnstiel T., Klahr H. \& Ercoano B. 2012. A\& A {\bf 539}: A148.

\bibitem[\protect\citeauthoryear{Bitsch et al.}{2013}]{B13} Bitsch B., Crida A., Morbidelli A., Kley W. \& Dobbs-Dixon I. 2013. A\& A {\bf 549}: A124.

\bibitem[\protect\citeauthoryear{Blaes \& Hawley}{1994}]{BH94} Blaes O.M. \& Hawley J.F., 1994, ApJ {\bf 421}: 163.

\bibitem[\protect\citeauthoryear{Brogi et al.}{2016}]{Brog16} Brogi M., Kok R.J., Albrecht S., Snellen I.A.G., Birkby J.L. \& Scharz H. 2016. ApJ {\bf 817}: 106.

\bibitem[\protect\citeauthoryear{Chambers}{2009}]{Cham09} Chambers J.E., 2009,ApJ {\bf 705}: 1206.

\bibitem[\protect\citeauthoryear{Chiang \& Goldreich}{1997}]{CG97} Chiang E.I. \& Goldreich P. 1997. ApJ {\bf 490}: 368.

\bibitem[\protect\citeauthoryear{Cieza et al.}{2015}]{Cie15} Cieza L., Williams J., Kourkchi E., Andrews S., Casassus S., Graves S. \& Schreiber M. 2015. arXiv:1504.06040.

\bibitem[\protect\citeauthoryear{Cleeves et al.}{2014}]{Cle14} Cleeves L.I., Bergin E.A., \& Adams F.C., 2014, ApJ {\bf 794}: 123.

\bibitem[\protect\citeauthoryear{Croll et al.}{2015}]{Crl15} Croll B., Albert L., Jayawardhana R., Cushing M., Moutou Cl. Lafreniere D., Johnson J.A., Bonomo Al.S., Deleuil M. \& Fortney J. 2015. ApJ {\bf 802}: 28.

%\bibitem[\protect\citeauthoryear{Coleman \& Nelson}{2014}]{CN14} Coleman G.A.L. \& Nelson R.P. 2014. MNRAS {\bf 445}: 479

\bibitem[\protect\citeauthoryear{D'Alessio et al.}{1998}]{Dal98} D'Alessio P., Cant${\rm \acute{o}}$ J., Calvet N. \& Lizano S. 1998. ApJ {\bf 500}: 411.

\bibitem[\protect\citeauthoryear{D'Alessio et al.}{1999}]{Dal99} D'Alessio P., Calvet N. \& Hartmann L. 1999. ApJ {\bf 527}: 893.

\bibitem[\protect\citeauthoryear{Dittkrist et al.}{2014}]{D14} Dittkrist K.-M., Mordasini C., Klahr H., Alibert Y. \& Henning T. 2014. A\& A {\bf 567}: A121.

\bibitem[\protect\citeauthoryear{Dullemond}{2012}]{RADMC} Dullemond C.P. 2012. ascl:1202.015.

\bibitem[\protect\citeauthoryear{Dullemond \& Dominik}{2005}]{DD05} Dullemond C.P. \& Dominik C. 2005. A\& A {\bf 434}: 971.

\bibitem[\protect\citeauthoryear{Fogel et al.}{2011}]{Fog11} Fogel J.K., Bethell T.J., Bergin E.A., Calvet N. \& Semenov D., 2011, ApJ {\bf 726}: 29.

\bibitem[\protect\citeauthoryear{Fortney et al.}{2005}]{For05} Fortney J.J., Marley M.S., Lodders K., Saumon D. \& Freedman R. 2005. ApJ {\bf 627}: L69-L72.

\bibitem[\protect\citeauthoryear{Fromang}{2013}]{Fro13} Fromang S. 2013. {\it Angular Momentum Transport During Star Formation and Evolution}. Hennebelle P. \& Charbonnel C. (eds). EAS Publications Series, {\bf 62}: 95.

\bibitem[\protect\citeauthoryear{Garaud et al.}{2013}]{Gara13} Garaud R., Meru F., Galvagni M. \& Olczak C. 2013. {\bf 764}: 146.

\bibitem[\protect\citeauthoryear{Gorti et al.}{2015}]{Gort15} Gorti U., Hollenbach D. \& Dullemond C. P. 2915. ApJ {\bf 804}: 29.

\bibitem[\protect\citeauthoryear{Gressel et al.}{2015a}]{Gres15a} Gressel O., Turner N.J., Nelson R.P. \& McNally C.P., 2015, ApJ {\bf 801}: 84.

\bibitem[\protect\citeauthoryear{Gressel et al.}{2015b}]{Gres15b} Gressel O. \& Pessah M.E., 2015, ApJ {\bf 810}: 59.

\bibitem[\protect\citeauthoryear{Haisch et al.}{2001}]{HLL01} Haisch K.E., Lada E.A. \& Lada C.J. 2001. ApJ {\bf 553}: L153.

\bibitem[\protect\citeauthoryear{Hasegawa \& Pudritz}{2010}]{HP10} Hasegawa Y. \& Pudritz R.E., 2010. ApJL {\bf 710}: L167.

\bibitem[\protect\citeauthoryear{Hasegawa \& Pudritz}{2011}]{HP11} Hasegawa Y. \& Pudritz R.E., 2011. MNRAS {\bf 417}: 1236.

\bibitem[\protect\citeauthoryear{Hasegawa \& Pudritz}{2012}]{HP12} Hasegawa Y. \& Pudritz R.E., 2012. ApJ {\bf 760}: 117.

\bibitem[\protect\citeauthoryear{Hasegawa \& Pudritz}{2013}]{HP13} Hasegawa Y. \& Pudritz R.E., 2013. ApJ {\bf 778}: 78.

\bibitem[\protect\citeauthoryear{Hayashi}{1961}]{Hay61} Hayashi C. 1961. PASJ {\bf 13}: 450.

\bibitem[\protect\citeauthoryear{Helling et al.}{2014}]{Hel14} Helling Ch., Woiteke P., Rimmer P., Kamp I., Thi W.-F. \& Meigerink R. 2014. {\it Life} {\bf 4}: 142.

\bibitem[\protect\citeauthoryear{Henning \& Semenov}{2013}]{SeHe13} Henning T. \& Semenov D. 2013. Chemical Reviews {\bf 113}: 9016-9042.

\bibitem[\protect\citeauthoryear{Hernandez et al.}{2007}]{Her07} Hernandez J., Calvet N., Briceno C., Hartmann L., Vivas A.K., Muzerolle J., Downes J., Allen L. \& Gutermuth R. 2007. ApJ {\bf 671}: 1784.

\bibitem[\protect\citeauthoryear{Ida \& Lin}{2004}]{IL04} Ida S. \& Lin D.N.C., 2004, ApJ {\bf 604}: 388.

\bibitem[\protect\citeauthoryear{Ida, Lin \& Nagasawa}{2013}]{ILN13} Ida S., Lin D.N.C. \& Nagasawa M., 2013, ApJ {\bf 775}: 42I.

\bibitem[\protect\citeauthoryear{Ikoma et al.}{2000}]{I00} Ikoma M., Nakazawa K. \& Emori H. 2000. ApJ {\bf 537}: 1013.

\bibitem[\protect\citeauthoryear{Jorgensen et al.}{2012}]{J12} Jorgensen J.K., Favre C., Bisschop S.E., Bourke T.L., van Dishoeck E.F. \& Schmalzl M. 2012. ApJL {\bf 757}: 4.

\bibitem[\protect\citeauthoryear{Kastner et al.}{2014}]{Kast14} Kastner J.H., Hily-Blant P., Rodriquez D.R., Punzi K. \& Forveille T. 2014. ApJ {\bf 793}: 55.

\bibitem[\protect\citeauthoryear{Kataria et al.}{2016}]{Kat16} Kataria T., Sing S.K., Lewis N.K., Visscher C., Showman A.P., Fortney J.J. \& Marley M.S. 2016. ApJ {\bf 821}: 9.

\bibitem[\protect\citeauthoryear{Kreidberg et al.}{2014}]{Kre14} Kreidberg L., Bean J.L., D\'esert J-M., Line M.R., Fortney J.J., Madhusudhan N., Stevenson K.B., Showman A.P., Charbonneau D., McCullough P.R., Seager S., Burrows A., Henry G.W., Williamson M., Kataria T. \& Homeier D. 2014. ApJL {\bf 793}: L27.

\bibitem[\protect\citeauthoryear{Kenyon \& Hartmann}{1987}]{KH87} Kenyon S.J. \& Hartmann L. 1087. ApJ {\bf 323}: 714.

\bibitem[\protect\citeauthoryear{Kley}{1999}]{K99} Kley W. 1999. MNRAS {\bf 303}: 696.

\bibitem[\protect\citeauthoryear{Kokubo \& Ida}{2002}]{KI02} Kokubo E. \& Ida S. 2002. ApJ {\bf 581}: 666.

\bibitem[\protect\citeauthoryear{Konopacky et al.}{2014}]{Kon14} Konospacky Q., Barman T., Macintosh B., Marous C. \& Savransky D. 2014. SPIE Newsroom: 10.1117/2.1201408.005546.

\bibitem[\protect\citeauthoryear{Kunz \& Balbus}{2004}]{KB04} Kunz M.W. \& Balbus S.A., 2004, MNRAS {\bf 348}: 355

\bibitem[\protect\citeauthoryear{Lee et al.}{2011}]{Lee11} Lee J.-M., Fletcher L.N. \& Irwin P.G.J., 2011, MNRAS {\bf 420}: 170.

\bibitem[\protect\citeauthoryear{Lee et al.}{2013}]{Lee13} Lee J.-M., Heng K., \& Irwin P.G.J., 2013, ApJ {\bf 778}: 97.

\bibitem[\protect\citeauthoryear{Lesniak \& Desch}{2011}]{LS11} Lesniak M.V. \& Desch S.J., 2011, ApJ {\bf 740}: 118.

\bibitem[\protect\citeauthoryear{Lin \& Papaloizou}{1993}]{LP93} Lin D.N.C. \& Papaloizou J.C.B. 1993. Protostars and Planets III. The University of Arizona Press.

\bibitem[\protect\citeauthoryear{Line et al.}{2011}]{Lin11} Line M.R., Vasisht G., Chen P., Angerhausen D. \& Yung Y.L., 2011, arXiv:1104:3183v2.

\bibitem[\protect\citeauthoryear{Line et al.}{2013}]{Lin13} Line M.R., Knutson H., Wolf A.S. \& Yung Y.L., 2013, ApJ {\bf 783}: 70.

\bibitem[\protect\citeauthoryear{Lynden-Bell \& Pringle}{1974}]{LB74} Lynden-Bell D. \& Pringle J.E. 1974 MNRAS {\bf 168}: 603.

\bibitem[\protect\citeauthoryear{Madhusudhan et al.}{2014}]{Mad14} Madhusudhan N. Amin M.A, \& Kennedy G.M., 2014, ApJ {\bf 794}: L12.

\bibitem[\protect\citeauthoryear{Madhusudhan \& Seager}{2009}]{MadSea09} Madhusudhan N. \& Seager S., 2009, ApJ {\bf 707}: 24.

\bibitem[\protect\citeauthoryear{Madhusudhan \& Seager}{2010}]{MadSea10} Madhusudhan N. \& Seager S., 2010, ApJ {\bf 725}: 261.

\bibitem[\protect\citeauthoryear{Mandell et al.}{2012}]{Mand12} Mandell A.M., Bast J., van Dishoeck E.F., Blake G.A., Salyk C., Mumma M.J. \& Villanueva G. 2012. ApJ {\bf 747}: 92.

%\bibitem[\protect\citeauthoryear{Marcus et al.}{2015}]{Mac15} Marcus P.S., Pei S., Jiang C-H. Barranco J.A. Hassanzadeh P. \& Leconanet D. 2015. ApJ {\bf 808}: 87.

\bibitem[\protect\citeauthoryear{Marley et al.}{1996}]{Mar96} Marley M.S., Saumon D., Guillot T., Freedman R.S., Hubbard W.B., Burrows A. \& Lunine J.I. 1996. Science {\bf 272}: 1919-1921.

\bibitem[\protect\citeauthoryear{Masset \& Papaloizou}{2003}]{MasPap03} Masset F. S. \& Papaloizou J.C.B. 2003. ApJ {\bf 588}: 494.

\bibitem[\protect\citeauthoryear{Masset et al.}{2006}]{M06} Masset F.S., Morbidelli A., Crida A. \& Ferreira J. 2006. {\bf 642}: 478.

\bibitem[\protect\citeauthoryear{Mathis et al.}{1977}]{Math77} Mathis J.S., Rumpl W. \& Nordsieck K.H. 1977. ApJ {\bf 217}: 425.

\bibitem[\protect\citeauthoryear{Matsumura et al.}{2009}]{SHT09} Matsumura S., Pudritz R.E. \& Thommes E.W. 2009. ApJ {\bf 691}: 1764.

\bibitem[\protect\citeauthoryear{McElroy et al.}{2013}]{McE03} McElroy D., Walsh C., Markwick A.J., Cordiner M.A., Smith K. \& Millar T.J. 2013. A\& A {\bf 550}: A36.

\bibitem[\protect\citeauthoryear{Miguel \& Kaltenegger}{2014}]{MigKal13} Miguel Y. \& Kaltenegger L., 2014, ApJ {\bf 780}: 166.

\bibitem[\protect\citeauthoryear{Montalto et al.}{2015}]{Mont15} Montalto M., Iro N., Santos N.C., Desidera S., Martins J.H.C., Figueira P. \& Alonso R. 2015. ApJ {\bf 811}: 55.

\bibitem[\protect\citeauthoryear{Mordasini et al.}{2012}]{Mord12} Mordasini C., Alibert Y., Georgy C., Dittkrist K.-M., Klahr H. \& Henning T. A\& A {\bf 547}: A112.

\bibitem[\protect\citeauthoryear{Moses et al.}{2013}]{Mos12} Moses J.I., Madhusudhan N., VisscherC. \& Freedman R.S. 2013. ApJ {\bf 763}: 25.

\bibitem[\protect\citeauthoryear{{\"O}berg et al.}{2007}]{O07} {\"O}berg K.I., Awad Z., Fraser H.J., Schlemmer S., van Dishoeck E.F. \& Linnartz H. 2007. ApJ {\bf 662}: L23.

\bibitem[\protect\citeauthoryear{{\"O}berg et al.}{2011}]{O11} {\"O}berg K.I., Murray-Clay R. \& Bergin E.A. 2011. ApJ {\bf 743}: L16.

\bibitem[\protect\citeauthoryear{Oklop{\v c}i{\'c} et al.}{2016}]{Okl16} Oklop{\v c}i{\'c} A., Hirata C.M. \& Heng K. 2016. arXiv.org:1605.07185.

\bibitem[\protect\citeauthoryear{Paardekooper et al.}{2011}]{P11} Paardekooper S.-J., Baruteau C., Crida A. \& Kley W. 2010. MNRAS {\bf 401}: 1950.

\bibitem[\protect\citeauthoryear{Paardekooper \& Terquem}{1999}]{PT99} Paardekooper S.-J. \& Terquem C., 1999. ApJ {\bf 521}: 823.

%\bibitem[\protect\citeauthoryear{Pignatale et al.}{2011}]{Pig11} Pignatale F.C., Maddison S.T., Taquet V., Brooks G. \& Liffman K. 2011. MNRAS {\bf 414}: 2386.

\bibitem[\protect\citeauthoryear{Pontoppidan et al.}{2014}]{Pon14} Pontoppidan K.M., Salyk C., Bergin E.A., Brittain S., Marty B., Mousis O. \& {\"O}berg K.L. 2014. Protostars and Planets VI. The University of Arizona Press.

\bibitem[\protect\citeauthoryear{Pudritz \& Norman}{1983}]{PN83} Pudritz R.E. \& Norman C.A., 1983, ApJ {\bf 274}: 677.

\bibitem[\protect\citeauthoryear{Pudritz et al.}{2007}]{PPV07} Pudritz R.E., Ouyed R., Fendt C. \& Brandenburg A. 2007. {\it Protostars and Planets V}. B. Reipurth, D. Jewitt, and K. Keil (eds.), University of Arizona Press, Tucson, 2006.

\bibitem[\protect\citeauthoryear{Punzi et al.}{2015}]{Pun15} Punzi K.M., Hily-Blant P., Kastner .H., Sacco G.G. \& Forveille T. 2015. ApJ {\bf 805}: 147.

\bibitem[\protect\citeauthoryear{Qi et al.}{2013a}]{Qi13} Qi C., {\"O}berg K.I. \& Wilner D.J. 2013. ApJ {\bf 765}: 34.

\bibitem[\protect\citeauthoryear{Qi et al.}{2013b}]{Qi13b} Qi C., {\"O}berg K.I., Wilner D.J., D'Alessio P., Bergin E., Andrews S.M., Blake G.A., Hogerheijde M.R. \& van Dishoeck E.F. 2013. Science {\bf 341}: 630.

%%\bibitem[\protect\citeauthoryear{Rafikov}{2005}]{Raf05} Rafikov R.R. 2005. ApJ {\bf 621}: 69R

%%\bibitem[\protect\citeauthoryear{Rafikov}{2011}]{Raf11} Rafikov R.R. 2011. ApJ {\bf 727}: 86R

%%\bibitem[\protect\citeauthoryear{Rice et al.}{2015}]{Rice15} Rice K., Lopez, E., Forgan D. \& Biller B. 2015. MNRAS {\bf 454}: 1940

\bibitem[\protect\citeauthoryear{Richert et al.}{2015}]{R15} Richert A.J.W., Feigelson E.D., Getman K.V. \& Kuhn M.A. 2015. arXiv:1508.01338.

\bibitem[\protect\citeauthoryear{Rowe et al.}{2014}]{Kep14} Rowe J.F., Bryson S.T., Marcy G.W, Lissauer J.J. et al. 2014. ApJ {\bf 784}: 45.

\bibitem[\protect\citeauthoryear{Seager et al.}{2010}]{Exo10} Seager S., editor. 2010. {\it Exoplanets}. pp 353-360. The University of Arizona Press.

\bibitem[\protect\citeauthoryear{Seager \& Deming}{2010}]{SD10} Seager S. \& Deming D. 2011. ARAA {\bf 48}:631.

\bibitem[\protect\citeauthoryear{Semenov et al.}{2010}]{Sem10} Semenov D., Hersant F., Wakelam V., Dutrey A., Chapillon E., Guilloteau St., Henning Th., Launhardt R., Pietu V. \& Schreyer. 2010. A\& A {\bf 522}: A42.

\bibitem[\protect\citeauthoryear{Semenov \& Wiebe}{2011}]{SW11} Semenov D. \& Wiebe D. 2011. ApJ {\bf 196}:25.

\bibitem[\protect\citeauthoryear{Shakura \& Sunyaev}{1973}]{SS73} Shakura N.I. \& Sunyaev R.A., 1973, A\& A {\bf 24}: 337.

\bibitem[\protect\citeauthoryear{Simon et al.}{2013}]{Simon13} Simon J.B., Bai X-N., Stone J.M., Armitage P.J. \& Beckwith K. 2013. {\bf 764}: 66.

\bibitem[\protect\citeauthoryear{Sing et al.}{2014}]{Sing14} Sing D.K., Wakeford H.R., Showman A.P., Nikolov N. et al. 2014. MNRAS {\bf 446}: 2428.

\bibitem[\protect\citeauthoryear{Smith et al.}{2004}]{Sea04} Smith I.W.M., Herbst E. \& Chang Q. 2004. MNRAS {\bf 350}: 323.

\bibitem[\protect\citeauthoryear{Snellen et al.}{2015}]{Snel15} Snellen I., de Kok R., Birkby J.L., Brandl B., Brogi M., Keller C., Kenworthy M., Schwarz H. \& Stuik R. 2015. A\& A {\bf 576}: A59. 

\bibitem[\protect\citeauthoryear{Stepinski}{1998}]{St98} Stepinski, T.F., 1998 Icarus {\bf 132}: 100.

\bibitem[\protect\citeauthoryear{Stevenson et al.}{2010}]{Stv10} Stevenson K.B. et al., 2010, Nature {\bf 464}: 1161.

\bibitem[\protect\citeauthoryear{Strom et al.}{1989}]{Strom89} Strom K.M., Strom S.E., Edwards S., Cabrit S. \& Skrutskie M.F. 1989. AJ {\bf 97}: 1451.

\bibitem[\protect\citeauthoryear{Thi et al.}{2013}]{Thi13} Thi W.-F., Kamp I., Woitke P., van der Plas G., Bertelsen R. \& Wiesenfeld L. 2013. A\& A {\bf 551}: 26.

\bibitem[\protect\citeauthoryear{Tielens}{2005}]{Tie05} Tielens A.G.G.M. 2005. {\it The Physics and Chemistry of the Interstellar Medium}. pp 85-115. Cambridge University Press.

\bibitem[\protect\citeauthoryear{Van Grootel et al.}{2014}]{VGro14} Van Grootel V et al., 2014, ApJ {\bf 786}: 2.

\bibitem[\protect\citeauthoryear{Venot et al.}{2014}]{Ven14} Venot O., Agundez M., Selsis F., Tessenyi M. \& Iro N., 2014, A\& A {\bf 562}: A51.

\bibitem[\protect\citeauthoryear{Walsh et al.}{2013}]{Wal13} Walsh C., Millar T.J. \& Nomura H. 2013. ApJL {\bf 766}: 23.

\bibitem[\protect\citeauthoryear{Walsh et al.}{2014}]{Wal14} Walsh C., Nomura H., Millar T.J. \& Aikawa Y. 2014. ApJ {\bf 747}: 114.

\bibitem[\protect\citeauthoryear{Ward}{1991}]{W91} Ward W.R. 1991. Horseshoe orbit drak. In {\it Lunar Planet. Sci. XXII}, pp. 1463-1464. Lunar and Planetary Institute, Houston.

\bibitem[\protect\citeauthoryear{Ward}{1997}]{W97} Ward W.R. 1997. ApJ {\bf 482}: L211.

\bibitem[\protect\citeauthoryear{Weingartner \& Draine}{2001}]{WD01} Weingartner, J.C., \& Draine, B.T. 2001. ApJ {\bf 548}: 296.

\bibitem[\protect\citeauthoryear{Wilking et al.}{1989}]{Wilk89} Wilking B.A., Lada C.J. \& Young E.T. 1989. ApJ {\bf 340}: 823.

%\bibitem[\protect\citeauthoryear{Willacy et al.}{2015}]{Wila15} Willacy K., Alexander C., Ali-Dib M., Ceccarelli C. , Charnley S.B., Doronin M., Ellinger Y., Gast P., Gibb E., Milam S. N., Mousis O., Pauzat F., Tornow C., Wirström E.S., Zicler E. \& Space Science Reviews (2015) pp. 1-40.

\bibitem[\protect\citeauthoryear{Williams et al.}{2013}]{Will13} Williams J.P., Cieza L.A., Andrews S.M., Coulson I.M., Barger A.J., Casey C.M., Chen C-C., Cowie L.L., Koss M., Lee N. \& Sanders D.B. 2013. arXiv:1307.7274.

\bibitem[\protect\citeauthoryear{Wilson et al.}{2015}]{Wil15} Wilson P.A., Sing D.K., Nikolov N., Lecavelier des Etangs A. et al. 2015. MNRAS {\bf 450}: 192.

\bibitem[\protect\citeauthoryear{Woitke et al.}{2009}]{Wot09} Woitke P., Kamp I. \& Thi W.-F. 2009. A\& A {\bf 501}: 383.

\bibitem[\protect\citeauthoryear{Woodall et al.}{2007}]{Wood07} Woodall J., Agundez M., Markwick-Kemper A.J. \& Millar T.J. 2007. A\& A {\bf 466}: 1197.

\bibitem[\protect\citeauthoryear{Wyttenback et al.}{2015}]{Wyt15} Wyttenback A., Ehrenreich D., Lovis C., Udry S. \& Pepe F. 2015. A\& A {\bf 577}: A62.

\bibitem[\protect\citeauthoryear{Zhang et al.}{2013}]{Z13} Zhang K., Pontoppidan K.M., Salyk C. \& Blake G.A. 2013. ApJ {\bf 766}: 82.

\bibitem[\protect\citeauthoryear{Zhang et al.}{2015}]{Zh15} Zhang K., Blake G.A. \& Bergin E.A. 2015. ApJ {\bf 806}: L7.

\bibitem[\protect\citeauthoryear{Zheng et al.}{2015}]{Z15} Zheng X., Kouwenhoven M.B.N. \& Wang L. 2015. arXiv:1508.01593.


\end{thebibliography}
\end{document}